\documentclass[12pt,preprint]{aastex}

\newcommand{\bdv}[1]{\mbox{\boldmath$#1$}}

\def\au{{\rm au}} 
 
\def\kms{{\rm km}\,{\rm s}^{-1}}
\def\masyr{{\rm mas}\,{\rm yr}^{-1}}
\def\kpc{{\rm kpc}}
\def\mas{{\rm mas}}

\def\muas{\mu{\rm as}}

\def\max{{\rm max}}

\def\rel{{\rm rel}}

\def\hel{{\rm hel}}
\def\geo{{\rm geo}}
\def\e{{\rm E}}
\def\bpi{{\bdv\pi}}
\def\bmu{{\bdv\mu}}

\def\btheta{{\bdv\theta}}

\def\bv{{\bf v}}

\begin{document}
\title{OGLE-2018-BLG-0532Lb: Cold Neptune With Possible Jovian Sibling}

\author{\textsc{
Yoon-Hyun Ryu$^{1}$, 
Andrzej Udalski$^{2}$,
Jennifer C. Yee$^{3}$, 
Matthew T. Penny$^{4}$,
Weicheng Zang$^{5}$,
\and
Michael D.Albrow$^{6}$, 
Sun-Ju Chung$^{1,7}$, 
Andrew Gould$^{8,4}$, 
Cheongho Han$^{9}$, 
Kyu-Ha Hwang$^{1}$, 
Youn Kil Jung$^{1}$, 
In-Gu Shin$^{1}$, 
Yossi Shvartzvald$^{10}$, 
Sang-Mok Cha$^{1,11}$, 
Dong-Jin Kim$^{1}$,
Hyoun-Woo Kim$^{1}$, 
Seung-Lee Kim$^{1,7}$, 
Chung-Uk Lee$^{1,7}$, 
Dong-Joo Lee$^{1}$,
Yongseok Lee$^{1,11}$, 
Byeong-Gon Park$^{1,7}$, 
Richard W. Pogge$^{4}$ \\
(KMTNet Collaboration)\\
Przemek Mr{\'o}z$^{2}$,
Micha{\l} K. Szyma{\'n}ski$^{2}$,
Jan Skowron$^{2}$,
Radek Poleski$^{4}$,
Igor Soszy{\'n}ski$^{2}$,  
Pawe{\l} Pietrukowicz$^{2}$,
Szymon Koz{\l}owski$^{2}$,
Krzysztof Ulaczyk$^{12}$,
Krzysztof A. Rybicki$^{2}$,
Patryk Iwanek$^{2}$,
Marcin Wrona$^{2}$\\
(OGLE Collaboration)\\
Shude Mao$^{5,13}$,
Pascal Fouqu\'e$^{14,15}$,
Wei Zhu$^{16}$,
Tianshu Wang$^{5}$\\
(CFHT microlensing collaboration)\\
} }

\affil{$^{1}$Korea Astronomy and Space Science Institute, Daejon
34055, Republic of Korea}

\affil{$^{2}$Warsaw University Observatory, Al.~Ujazdowskie~4, 00-478~Warszawa,
Poland}

\affil{$^{3}$ Center for Astrophysics $|$ Harvard \& Smithsonian, 60 Garden
St., Cambridge, MA 02138, USA}

\affil{$^{4}$Department of Astronomy, Ohio State University, 140 W.
18th Ave., Columbus, OH 43210, USA}

\affil{$^{5}$Physics Department and Tsinghua Centre for
Astrophysics, Tsinghua University, Beijing 100084, China}

\affil{$^{6}$University of Canterbury, Department of Physics and
Astronomy, Private Bag 4800, Christchurch 8020, New Zealand}

\affil{$^{7}$Korea University of Science and Technology, Korea, 
(UST), 217 Gajeong-ro, Yuseong-gu, Daejeon, 34113, Republic of Korea}

\affil{$^{8}$Max-Planck-Institute for Astronomy, K\"{o}nigstuhl 17,
69117 Heidelberg, Germany}

\affil{$^{9}$Department of Physics, Chungbuk National University,
Cheongju 28644, Republic of Korea}

\affil{$^{10}$IPAC, Mail Code 100-22, Caltech, 1200 E. California Blvd., 
Pasadena, CA 91125, USA}


\affil{$^{11}$School of Space Research, Kyung Hee University,
Yongin, Kyeonggi 17104, Republic of Korea}

\affil{$^{12}$Department of Physics, University of Warwick, Gibbet Hill Road,
Coventry, CV4~7AL,~UK}

\affil{$^{13}$National Astronomical Observatories, Chinese Academy of Sciences, A20 Datun Rd., Chaoyang District, Beijing 100012, China}

\affil{$^{14}$CFHT Corporation, 65-1238 Mamalahoa Hwy, Kamuela, Hawaii 96743, USA}

\affil{$^{15}$Universit\'e de Toulouse, UPS-OMP, IRAP, Toulouse, France}

\affil{$^{16}$Canadian Institute for Theoretical Astrophysics, University of Toronto, 60 St George Street, Toronto, ON M5S 3H8, Canada}

\begin{abstract}

We report the discovery of the planet OGLE-2018-BLG-0532Lb, with very obvious
signatures in the light curve that lead to an estimate of the planet-host
mass ratio $q=M_{\rm planet}/M_{\rm host}\simeq 1\times10^{-4}$.  Although there
are no obvious systematic residuals to this double-lens/single-source
(2L1S) fit, we find that $\chi^2$ can be significantly improved by
adding either a third lens (3L1S, $\Delta\chi^2=81$) 
or second source (2L2S, $\Delta\chi^2=65$) to the
lens-source geometry.  After thorough investigation, we conclude that
we cannot decisively distinguish between these two scenarios and 
therefore focus on the robustly-detected planet.  
However, given the possible presence of a second planet,
we investigate to what degree and with what probability such additional
planets may affect seemingly single-planet light curves.  Our
best estimates for the properties of the lens star and the secure planet are: a host mass $M\sim 0.25\,M_\odot$, 
system distance $D_L\sim 1\,\kpc$ and planet mass
$m_{p,1}= 8\,M_\oplus$ with projected separation $a_{1,\perp}=1.4\,\au$.
However, there is a relatively bright $I=18.6$ (and also relatively blue)
star projected within $<50\,\mas$ of
the lens, and if future high-resolution images show that this is
coincident with the lens, then it is possible that it is the lens,
in which case, the lens would be both more massive and more distant
than the best-estimated values above.

\end{abstract}

\keywords{gravitational lensing: micro; planetary systems}

\section{{Introduction}
\label{sec:intro}}

Based on the second detection of a Neptune-class planet beyond the
snow line, \citet{ob05169} had already suggested that such ``cold Neptunes''
are the most common type of planet.  As microlensing-planet discoveries
continued to accumulate and populate the $(\log q,\log s)$ diagram
(Figure~7 from \citealt{ob160596}), it became manifest that cold
Neptunes are at least more numerous than cold planets of greater mass.
Here, $q$ is the planet-host mass ratio and $s$ is the planet-host
projected separation normalized to the Einstein radius $\theta_\e$.
These $(\log q,\log s)$ diagrams show that planet detections are roughly
uniform over $-4.3< \log q < -2$.  Because higher-mass planets have
larger caustics, and so larger cross sections for detection, a uniform
rate of detection implies more planets at lower mass.

The same $(\log q,\log s)$ diagrams show a sharp cut off in detections
at $\log q\simeq -4.3$, i.e., $q\simeq 5\times 10^{-5}$.  
See Figure~3 of \citet{ob171434}.  This could
in principle reflect a sharp cutoff in the existence of cold planets
at lower masses, but it also might simply reflect a cutoff in sensitivity
of present-day microlensing experiments.

However, two studies concluded that the first explanation is correct:
cold planet frequency reaches a peak at cold Neptunes and then
declines toward lower masses.  First, \citet{suzuki16} studied planets
detected by MOA and compared these to the planet sensitivity of the MOA
sample, as judged by a $\Delta\chi^2$ criterion.  They concluded that
the cold-planet frequency (as a function of $\log q$) peaks around
$\log q\sim -4$.  Second, \citet{ob171434} studied the complete
sample of seven microlensing planets with well-measured mass 
ratios $\log q<-4$.  They developed a new ``$V/V_\max$'' method that
is logically independent of the \citet{suzuki16} method.  The data
samples were also largely independent.  \citet{ob171434} concluded
that if the mass-ratio function in this low-mass-ratio regime is
modeled as a single power law, then it
was falling toward lower $\log q$ in this range,
thus confirming the results of \citet{suzuki16}.  In particular,
\citet{ob171434} found that OGLE-2017-BLG-1434Lb would have been detected
(and been well characterized) even if it had been 30 times lighter
than it actually is, i.e., $\log q\simeq -5.7$.

\citet{kepq} derived the planet mass-ratio function of 
{\it Kepler} transiting planets by combining the planet-radius distribution 
and the empirical mass-radius relation. They found a break at 
$\log{q} \approx -4.5$. A similar break (or pile-up) is also found by 
\citet{wu18}, 
who inferred the Kepler planet masses from the photo-evaporation 
gap. Because microlensing planets are typically outside the snow line by 
a factor two or more, while {\it Kepler} transiting planets are inside the 
snow line by a similar factor, these results enable one to study whether 
(and how) the break in the planet mass-ratio function is strongly 
influenced by the presence (or absence) of icy material in the 
proto-planetary disk.

\citet{kb170165} subsequently analyzed the 15 planets with
well-determined mass ratios $q<3\times 10^{-4}$ and concluded from
their approximately uniform distribution in $\log q$ that the break
in the microlensing mass-ratio function is both quite low,
$q_{\rm break}\simeq 5.6\times 10^{-5}$
(i.e., $\log q_{\rm break}\simeq -4.25$), and quite severe, to the
point of approximating a cut off.  They also pointed toward possible
evidence for a ``pile up'', at or just above the break.   This
break point would be more consistent with the {\it Kepler} break
found by \citet{kepq}, although probably much more severe.

Additional detections of planets in this low-mass regime will
be crucial for resolving the position and severity of the break
in the microlensing mass-ratio function.  
Here we analyze OGLE-2018-BLG-0532, whose lensing system contains
at least one planet, a cold Neptune with mass ratio
$q=(1.0\pm 0.2)\times 10^{-4}$, i.e., right at the somewhat arbitrary
``low mass-ratio'' boundary of \citet{ob171434} and near the upper
end of the possible ``pile-up'' noted by \citet{kb170165}.
We note that \citet{ob151670}
have reported a planet that also straddles this
boundary, with $q=(1.00\pm 0.17)\times 10^{-4}$.  That is, both
of these planets would meet the sample conditions of the
\cite{kb170165} study.

Moreover, OGLE-2018-BLG-0532 also contains evidence
for a second planet,  a Jovian-class body that would lie 
at a projected distance that is either 2.65 times greater or 2.65 times smaller
than the Neptune mass-ratio planet. Fitting for the additional planet results in an improvement of
$\Delta\chi^2=81$, which is formally very significant. At the same time, the candidate planet 
does not result in any recognizable signatures, either in the 
original light curve or in the residuals to the single-planet fit.

\citet{shin-3L1S} studied the problem of such weak third-body signatures
in apparently single-planet microlensing events.  At the time of their
study (and still today), there were only two cases of two-planet
microlensing systems with unambiguous signatures for both planets:
OGLE-2006-BLG-109 \citep{ob06109,ob06109b} and OGLE-2012-BLG-0026
\citep{ob120026}. In both cases, the underlying events had high
magnifications, (normalized impact parameters $u_0=0.0035$ and $u_0=0.0095$, 
respectively),
which made them especially sensitive to planets
via their ``central caustics'' \citep{griest98}, and thus in particular
makes them sensitive to multiple planets in the same system
\citep{gaudi-multi}.  In fact, both systems followed the ``factoring''
(or ``superposition'')
of caustic signatures already predicted by \citet{planet-factor}.

However, \citet{shin-3L1S} considered that if there were additional planets
in a given system with one detected planet, it was at least as likely
that they would give rise to weak signatures that may have escaped
notice as to strong signatures that were obvious, at least in the
residuals to the single-planet fit.  This led them to search for
additional bodies in eight apparently single-planet systems
in high or moderately high magnification events,
OGLE-2005-BLG-071 \citep{ob05071},
OGLE-2005-BLG-169 \citep{ob05169},
MOA-2007-BLG-400 \citep{mb07400},
MOA-2008-BLG-310 \citep{mb08310},
MOA-2009-BLG-319 \citep{mb09319},
MOA-2009-BLG-387 \citep{mb09387},
MOA-2010-BLG-477 \citep{mb10477}, and
MOA-2011-BLG-293 \citep{mb11293}.
These have, respectively, normalized impact parameters
$u_0 = (
0.023,     
0.00012,   
0.0005,    
0.0030     
0.0062,    
0.08,      
0.0034,    
0.0035     
)$.

\citet{shin-3L1S} found that allowing for a third body led to
$\Delta\chi^2$ improvements of 143, 78, and 50
for events MOA-2009-BLG-319, MOA-2008-BLG-310, and MOA-2010-BLG-477,
respectively.  All other events had $\Delta\chi^2<30$.  Given that
none of these events met their adopted threshold of $\Delta\chi^2>500$,
they only set upper limits on additional planets in all cases.  They
showed only one two-planet fit, i.e., for MOA-2009-BLG-319, which had
the highest $\Delta\chi^2$.  As we will show is also the case for
OGLE-2018-BLG-0532, the single-planet fit did not display any particularly
noticeable residuals, and the two-planet fit did not result in any
obvious improvements.

Subsequently, \citet{ob141722} showed that the OGLE-2014-BLG-1722
light curve has two unambiguous deviations from a \citet{pac86} fit that
are well explained by a 3L1S model with two planets.  One deviation is
a dip and so can only be due to a planet, but the other is a smooth
bump, which could also in principle be caused by a second source.
This 2L2S model is disfavored by only $\Delta\chi^2=5.7$, but after
taking consideration of auxiliary information, \citet{ob141722} concluded
that the 3L1S model is favored by 3.1 sigma and so consider that this
is the ``likely'' interpretation.  Hence, this is currently the best
case for a relatively weak detection of a second planet, although the 
3L1S interpretation remains less than absolutely secure.

For OGLE-2018-BLG-0532, the smooth character of the residuals from the
2L1S fit also imply that one should test for 2L2S (in addition to 3L1S)
solutions.  Although we find that 3L1S is preferred by $\Delta\chi^2=16$ relative to 2L2S,
we consider that the evidence is too weak to claim the existence of
a second planet.  Nevertheless, the possibility of such a planet prompts
us to investigate how such ``weak detections'' are likely to appear
in the data.

\section{{Observations}
\label{sec:obs}}

OGLE-2018-BLG-0532 is at (RA,Dec) = (17:59:56.02,$-28$:59:51.9)
corresponding to $(l,b)=(1.54,-2.73)$.  It was announced to the
community by the Optical Gravitational Lensing Experiment 
(OGLE, \citealt{ogle-iv})
Early Warning System (EWS, \citealt{ews1,ews2})
at UT 13:54 on 9 April 2018.
OGLE observed this field (BLG505) in 2018 at a cadence
$\Gamma=1\,{\rm hr}^{-1}$ from its 1.3m telescope with $1.4\,{\rm deg}^2$
camera at Las Campanas Observatory in Chile.  These data show a clear
anomaly, but are adversely impacted by long weather gaps during the
anomaly.

The Korea Microlensing Telescope Network
(KMTNet, \citealt{kmtnet}) also observed this event from its
three identical 1.6m telescopes at CTIO (Chile, KMTC), 
SAAO (South Africa, KMTS) and SSO (Australia, KMTA), each equipped
with identical $4\,{\rm deg}^2$ cameras.  The event lies in three
overlapping KMTNet fields, BLG03, and BLG43, with a combined
cadence of $\Gamma = 4\,{\rm hr^{-1}}$ from KMTC and 
$\Gamma = 6\,{\rm hr^{-1}}$ from KMTS and KMTA.  In fact, there
are some observations from KMTC BLG04 as well, with a cadence
of $\Gamma = 1\,{\rm hr^{-1}}$.  The event was rediscovered at the
end of the season by the KMTNet eventfinder \citep{eventfinder}
as KMT-2018-BLG-1161.

The great majority of observations were carried out in the $I$ band 
with occasional $V$-band observations made
solely to determine source colors.
All reductions for the light curve
analysis were conducted using variants of image subtraction
\citep{alard98}, either DIA \citep{wozniak2000} or pySIS \citep{albrow09}.

Note that the errors were renormalized so that the $\chi^2$ per degree
of freedom (dof) at each observatory is approximately equal to unity.
In fact, the careful reader will note that the best model has a total
$\chi^2/{\rm dof}= 0.997$.  This is because the models slightly
improved after the initial renormalization, but we did not the repeat
renormalization and fitting (as is sometimes done) 
due to the exceptionally heavy computational load.

\section{{Analysis}
\label{sec:analysis}}

\subsection{{Search for 2L1S Solutions}
\label{sec:2L1S}}

\begin{figure}
\plotone{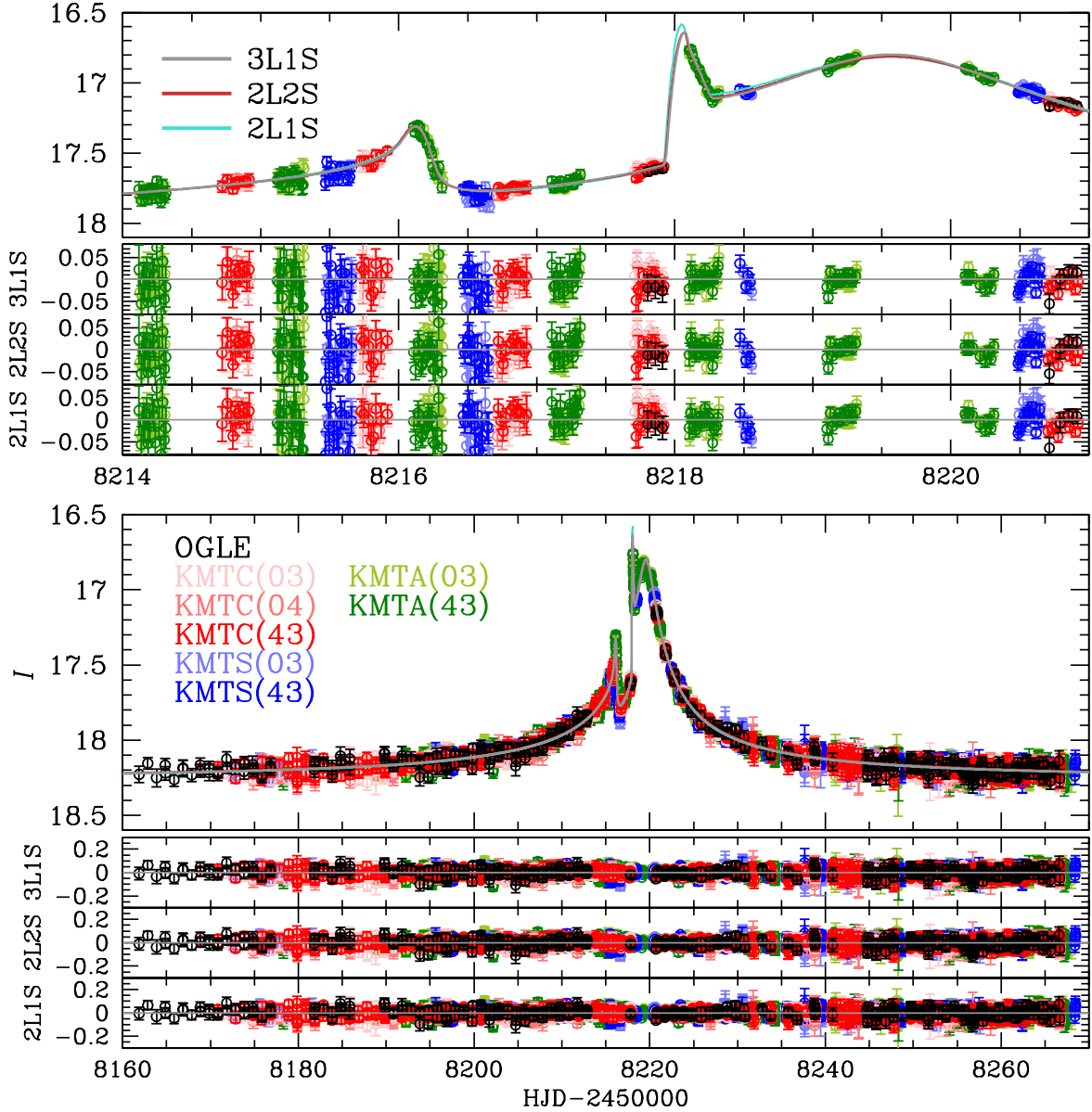}
\caption{Light curve and best-fit 3L1S (star plus two planets)
model of OGLE-2018-BLG-0532, with a zoom of the perturbation region
in the upper set of three panels, which also shows the best-fit 2L2S
and 2L1S models. The other degenerate models from 
Table~\ref{tab:3L1S}, ~\ref{tab:2L2S}, and ~\ref{tab:2L1S} 
are indistinguishable from these.
The broad flat trough lasting $\sim 1.5$ days, together with the adjacent
peaks is due to a mass ratio $q=1\times 10^{-4}$ planet. The second planet,
which is about 25 times more massive, does not give rise to obvious
signatures.  The panels immediately below the light-curve panels show
the residuals to this fit, while
the bottom panels show residuals to the 2L2S and 2L1S fits.
The differences between these panels are not obvious
to the eye. The 3L1S fit is favored by $\Delta\chi^2=81$ relative to the
2L1S fit and by $\Delta\chi^2=16$ relative to the 2L2S fit.
}
\label{fig:lc}
\end{figure}

Figure~\ref{fig:lc} shows the OGLE and KMTNet data, together with
the final best-fit triple-lens/single-source (3L1S) model.  
The duration of noticeable magnification is
quite long, $\ga 100\,$days, and the peak of the main event is relatively sharp.
Together, these characteristics would normally indicate a high
magnification event, $A_\max\gg 1$.  However, the flux at the 
peak is only a factor
$F_{\rm peak}/F_{\rm base}\sim 3.6$ above baseline.  These two indicators
can only be reconciled if the event is heavily blended, $f_b\gg f_s$,
where $f_s$ and $f_b$ are the source flux and the blended flux, respectively,
and where $F_{\rm base} = f_s + f_b$.

The major anomaly, which is shown in the upper (zoomed) panel
of Figure~\ref{fig:lc}, is comprised of
a trough that lasts $\sim 1.5$ days, flanked by two spikes, with the
post-trough spike being particularly sharp.  Such troughs can only
be produced by a minor-image perturbation.  The unperturbed light curve
is produced by two images that, according to Fermat's principle,
occur at extrema of the time-delay surface.  The larger (major) image, on one
side of the primary lens, is at a local minimum of this surface and is 
extremely stable.  The smaller (minor) image, on the opposite side, is at
a saddle point and is highly unstable.  Hence, it can be virtually
annihilated by a planet sitting at or near the position of the minor
image.  For high-magnification events $A_++A_-=A\gg 1$, the ratio of these
magnifications $A_+/A_-=(A+1)/(A-1)\rightarrow 1$.  Hence, up to half
the light can be briefly eliminated as a planet passes near to this
image.  On the flanking sides of such troughs are two caustics (lines
of formally infinite magnification), which continue as extended
ridges in magnification beyond the finite extent of the caustics 
themselves.  Thus, the form of the observed perturbation is exactly
as expected for such a geometry, and there are no other known geometries
that can generate this light-curve morphology.

Notwithstanding this qualitative analysis, we begin by conducting a systematic
grid search for 2L1S models over a range of $(s,q)$ geometries.
We hold these two parameters
fixed at the grid points, while five others 
$(t_0,u_0,t_\e,\rho,\alpha)$ are allowed to vary in a Monte Carlo 
Markov chain (MCMC).  The first three \citep{pac86} parameters
$(t_0,u_0,t_\e)$ are seeded at values derived from a 1L1S fit
to the light curve with the anomaly removed.  The angle $\alpha$
between the binary axis and lens-source relative proper motion $\bmu_\rel$ 
is seeded at six equally spaced
points around the unit circle, while 
$\rho=\theta_*/\theta_\e$, i.e., the source radius normalized to the
Einstein radius, is seeded at $\rho=5\times 10^{-4}$. As usual, there
are two flux parameters $(f_s,f_b)_i$ for each observatory $i$, so that
the predicted flux is given by 
$F_i(t) = f_{s,i}A(t;s,q,\alpha,t_0,u_0,t_\e,\rho) + f_{b,i}$.  
During the brief intervals when the source passes over or very close
to the caustic, it is partially resolved by the caustic, and so the
source brightness profile becomes relevant.  We characterize this
by a linear limb-darkening coefficient $\Gamma_I=0.50$
\citep{claret00}, in accordance with the source-type determination
of Section~\ref{sec:cmd}.

As anticipated, we find only one 2L1S solution, 
the parameters for which are given in Table~\ref{tab:2L1S}.  Before continuing,
we remark on the extreme level of blending $f_b/f_s \sim 40$.  That is,
despite the fact that the baseline appears relatively bright, 
$I_{\rm base}=18.25$ in the roughly-calibrated OGLE-IV photometry
(see Section~\ref{sec:photometry}), the source
is extremely faint, $I_s\sim 22.25$.  

\begin{figure}
\plotone{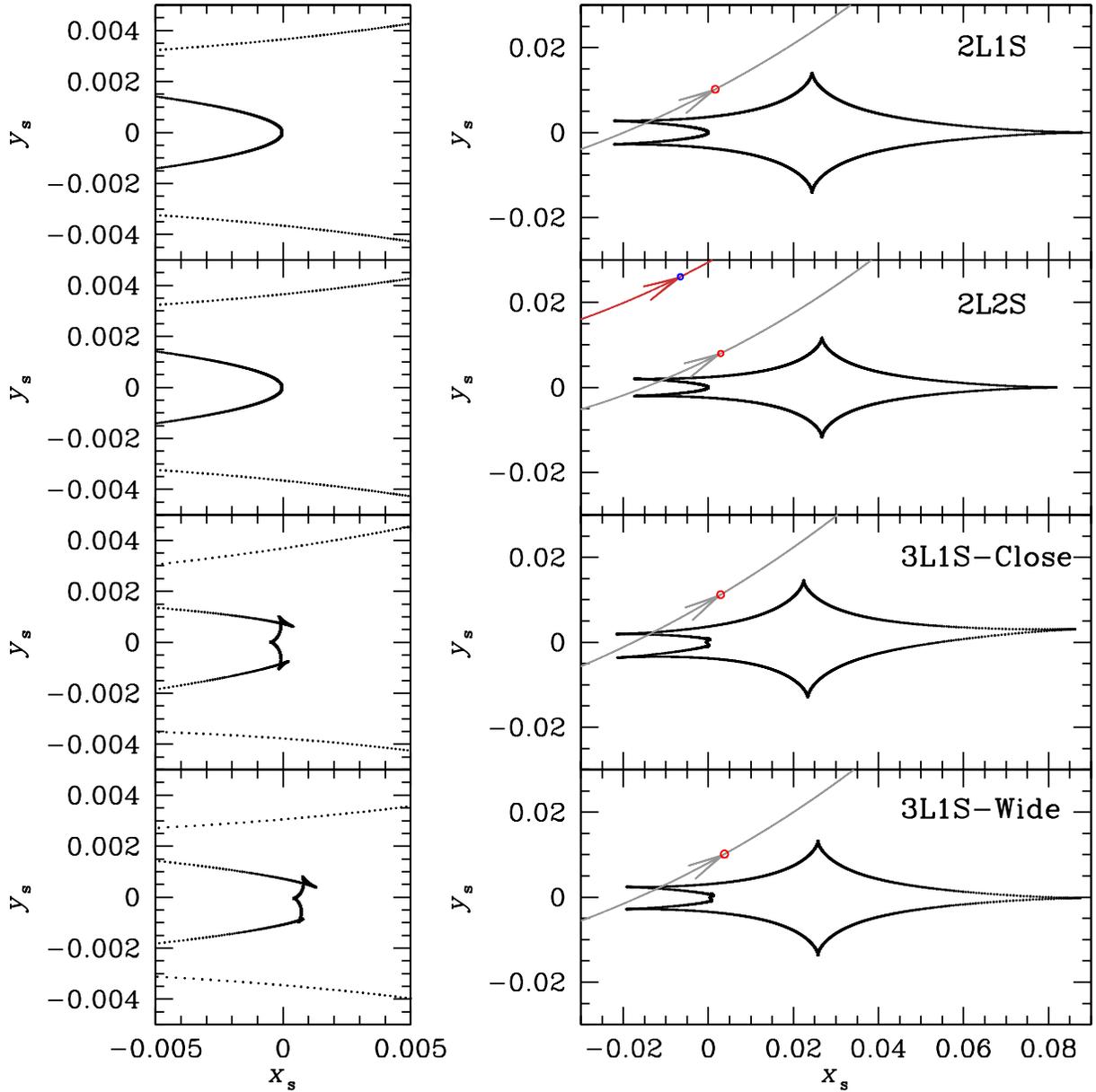}
\caption{Caustic geometries for the 
2L1S (Section~\ref{sec:2L1S}),
2L2S (Section~\ref{sec:2L2S}), and
3L1S (Section~\ref{sec:3L1S}) solutions.  The left panels show zooms of
these geometries in the neighborhood of the host.  In all cases, the
main feature, a pronounced dip near the peak, is caused by the
source transiting the magnification trough at the ``back end'' of a resonant
caustic.  In the 2L2S solution, there is a second source (red line)
passing well away from the caustic that accounts for the residuals.
The two sources are shown at the same time.  In the bottom two panels,
a second planet (3L1S) distorts the caustic near the central cusp
(left panels), which then accounts somewhat better for the residuals
to the 2L1S fit.
}
\label{fig:caustic}
\end{figure}

Figure~\ref{fig:caustic} shows the caustic geometry for the 2L1S
solution presented here, as well as for additional solutions
presented in Sections~\ref{sec:3L1S} and \ref{sec:2L2S}.  It shows
that the pronounced dip near the peak is due to the source transiting
the ``back end'' of a resonant caustic.

\subsection{{2L1S Solutions with Parallax}
\label{sec:2L1S_parallax}}

In spite of the faintness of the source (hence, high fractional errors
$\sigma(F)/f_s$), the long Einstein timescale $t_\e\sim 120\,$days 
implies that the microlens parallax may well be measurable.  
In particular, \citet{smp03} showed that the strength of the parallax
signal basically scales as $t_\e^4$.  This introduces two additional
parameters $(\pi_{\e,N},\pi_{\e,E})$, i.e., the components in equatorial 
coordinates of the ``vector microlens parallax'' $\bpi_\e$ 
\citep{gould92,gould00},
\begin{equation}
\bpi_\e = {\pi_\rel\over\theta_\e}\,{\bmu_\rel\over\mu_\rel},
\label{eqn:pie}
\end{equation}
where $\theta_\e$ is the Einstein radius,
\begin{equation}
\theta_\e = \sqrt{\kappa M\pi_\rel};
\qquad
\kappa\equiv {4G\over c^2\,\au} \simeq 8.14\,{\mas\over M_\odot},
\label{eqn:thetae}
\end{equation}
$M$ is the total mass of the lens, and $\pi_\rel$ is the
relative lens-source parallax.

Because the parallax effect (due to Earth's
orbital motion) can be degenerate with orbital motion of the lens
\citep{mb09387,ob09020}, one must also introduce two additional
parameters representing linearized orbital motion 
in order to test for such correlations.  We
adopt $d\alpha/dt$ and $ds/dt$ for the time rates of change of the binary's 
orientation and separation, respectively.  However, we find no significant
correlation between the orbital motion and parallax parameters.  Instead,
the orbital parameters become ``attracted'' to solutions that model
minor fluctuations in the 2017 (i.e., baseline) data as being due to
a planetary caustic that fortuitously ``orbited'' to the source position
during 2017.  This leads to small $\chi^2$ ``improvements'' that are
entirely spurious.  Because the orbital-motion parameters are not
in fact correlated with the parallax parameters, which was the original
reason for introducing them, we suppress the orbital-motion parameters.

For sources lying near the ecliptic, which includes essentially all
microlensing toward the Galactic bulge, one must check for the 
``ecliptic degeneracy'', which approximately takes
$(u_0,\pi_\e,\alpha,d\alpha/dt) \rightarrow -(u_0,\pi_\e,\alpha,d\alpha/dt)$.
We indeed find two such  solutions with very similar $\chi^2$, which
are both shown in Table~\ref{tab:2L1S}.  We also search for jerk-parallax
solutions \citep{gould04}, which would approximately take 
$\pi_{\e,N}\rightarrow -\pi_{\e,N}$, while leaving the other parameters 
approximately unchanged, but this search does not lead to viable solutions.

We note that most parameters are fairly similar between the ``parallax''
and ``standard'' models, except of course the parallax, which is
newly introduced.  The $\Delta\chi^2=71$ is highly significant, and
corresponding to this, the parallax amplitude $\pi_\e$ is measured
to about 13\%.  By far the largest change, apart from the parallax,
is the mass ratio, which drops by roughly 25\% in the parallax solutions.

One point of possible concern is that $\pi_{\e,E}$ is very close to
zero, with very small measurement errors, while $\pi_{\e,N}$ is much
larger.  This might be regarded as worrisome because 
Earth's acceleration is essentially East, and this induces an asymmetry
in the light curve, which is quite easy to measure because it is not
strongly correlated with any other parameters.  Hence, if the vector parallax
were actually close to zero, $\bpi_\e\sim 0$, then this would be robustly
reflected in $\pi_{\e,\parallel}\simeq\pi_{\e,E}$ but might be masked by
subtle long-term systematics in the light curve that gave rise to
a spurious $\pi_{\e,\perp}\simeq\pi_{\e,N}$.  

However, first, it will be straightforward to show that the parallax
must obey $\pi_\e\ga 0.2$ (see Section \ref{sec:blend}).  We do not discuss these arguments in 
detail here to avoid repeating them later.  Second, while $|\pi_{\e,E}|$
is small compared to $|\pi_{\e,N}|$, it is strongly inconsistent with 
zero, and quite consistent with the lower limit on $\pi_\e$ just
given.  Therefore, at this point there is no reason to doubt
the parallax measurement.  Nevertheless, we will return in Section \ref{sec:blend}
to the question of whether it could be several $\sigma$ lower than the
values reported in Table~\ref{tab:2L1S}.

\subsection{{Decision to Investigate 3L1S Solutions}
\label{sec:2L1Sprob}}


The residuals to the 2L1S model shown in Figure~\ref{fig:lc} do
not exhibit pronounced deviations, and therefore one would not
under normal circumstances search for a third body.  Our decision
to carry out such a 3L1S investigation was prompted by ``accidental''
developments in the course of the 2L1S investigation, i.e., 
apparent ``problems'' in the 2L1S solution that were all eventually resolved.
For completeness, we describe this process in the Appendix, but we
do not divert the reader's attention with the details here.

\subsection{{Search for 3L1S Solutions}
\label{sec:3L1S}}

We begin by considering static 3L1S models with 10 geometric parameters, i.e.,
the seven ``standard'' 2L1S parameters (with $(q,s)$ renamed $(q_1,s_1)$), plus
three additional parameters, $(q_2,s_2,\psi)$.  These are, respectively,
the mass ratio of second companion to the primary, the normalized 
separation between these, and the angle between primary-secondary
and primary-tertiary directions.  We conduct this search by simply
adding a new component with $q_2=0.3$ and $s_2=0.05$ and at several different
values of $\psi$ and then using MCMC to search for a local minimum.  
Such $s_2\ll 1$ models are in the extreme Chang-Refsdal
regime, in which the pseudo-shear\footnote{Strictly speaking, the shear
refers to the wide regime $s_2\gg 1$, where it is $\gamma_2^w = q_2/s_2^2$.
Based on the close-wide degeneracy, model pairs with 
$\gamma_2^c\sim\gamma_2^w$ should produce similar light curves.  For 
simplicity, we call therefore call $\gamma_2^c$ the ``pseudo-shear''.}
$\gamma_2^c \equiv q_2 s_2^2$ is an approximate
invariant.  Therefore, one expects that (if the choice of $\psi$ 
approximately conforms to the physical configuration
of the triple lens) the $\gamma_2^c$ parameter-combination will quickly
approach the true value in the MCMC, while the subsequent
disambiguation between $q_2$ and $s_2$ at fixed $\gamma_2^c$ will proceed
much more slowly.  In fact, we found that the first expectation
was confirmed but the second was not.  Instead, the MCMC became ``stuck''
and could not proceed toward the $(q_2,s_2)$ minimum, probably due
to the weakness of the $\chi^2$ gradient.  We therefore
substituted $s_2\rightarrow \gamma_2^c$ as an MCMC variable, after which
the minimum was found relatively quickly.

This search yields solutions with $q_2\ll 1$ and $s_2<0.5$.  Hence, one
expects a similar model light curve from the 
corresponding wide solution
$(q_2,s_2)\rightarrow(q_2,s_2^{-1})$ \citep{griest98,ob120026,song14}.
We find that this yields a model with very similar $\chi^2$.

We then add the two parallax parameters $(\pi_{\e,N},\pi_{\e,E})$.
After finding the best fit, we attempt to add orbital motion
(to first one, then the other companion).  However, as with the 2L1S
solutions, the $\chi^2$  only improves when the planetary caustic
becomes ``attached'' to noise features in the baseline.  Hence,
we again remove the orbital-motion parameters.

The best fit among these four solutions (close,wide)$\times(u_0<0,u_0>0)$ 
is shown in Figure~\ref{fig:lc},
together with the residuals of the data from this model.  By eye, the
improvement relative to the 2L1S residuals looks modest.

The four solutions
are shown in Table~\ref{tab:3L1S}.
Comparison of this table with Table~\ref{tab:2L1S} shows that the 
3L1S and 2L1S solutions are overall very similar (except for the addition
of three parameters due to the second planet).  This additional planet
is about 25 times more massive than the first planet, which is qualitatively
similar to the ratio of the masses of Jupiter and Neptune.  The addition
of this planet improves the fit by $\Delta\chi^2=81$.  This is formally
very significant, but it would be by far the lowest $\Delta\chi^2$
to justify an ``officially accepted'' planet.  Therefore, some discussion
is required to assess the reality of this putative planet.  See 
Section~\ref{sec:discuss}.

\subsection{{Search for 2L2S Solutions}
\label{sec:2L2S}}

As mentioned in Section~\ref{sec:2L1Sprob}, there are no obvious systematic
residuals to the 2L1S solution.  This means that the evidence for the
third body is not an additional caustic crossing 
(unlike e.g., OGLE-2006-BLG-109 and OGLE-2012-BLG-0026).  Whenever the
evidence for an additional body lacks such caustic features, one must
always test whether it can be generated by an additional source
rather than an additional lens, i.e., in this case 2L2S.  We therefore
begin by adding three parameters to the 2L1S geometry, which add a second
source with fixed projected offset from the first source.  That is, we
rename $(t_0,u_0)\rightarrow (t_{0,1},u_{0,1})$, introduce a second pair
$(t_{0,2},u_{0,2})$, and then also introduce the flux ratio of the second
source to the first in $I$ band, $q_{F,I}$.  This yields a substantial
improvement $\Delta\chi^2=124$ relative to the standard 2L1S solution.
We then add the two parallax parameters, $(\pi_{\e,N},\pi_{\e,E})$,
which yields an additional $\Delta\chi^2=12$ improvement.  Normally,
such a modest improvement would not be considered sufficient for a
reliable parallax measurement.  However, in the present case, we
must include these parameters in order to allow a fair comparison
with the 3L1S solutions.  Comparison of Tables~\ref{tab:2L2S} and \ref{tab:3L1S}
shows that the 3L1S solution is preferred by $\Delta\chi^2=16$
relative to the 2L2S solution.

We note that the microlensing parameters of the 3L1S and 2L2S
parallax solutions are qualitatively similar, but with some
quantitative differences.  The most notable differences are that
the mass ratio of the robustly detected planet and the (first) source
radius are smaller by a factor $\simeq 0.7$ in the 2L2S solution.  This mainly reflects
the fact that in high-magnification events, $t_*\equiv \rho t_\e$
and $q t_\e$ are approximate invariants \citep{mb11293}, and the value
of $t_\e$ is substantially higher in the 2L2S solution.  Note in
particular that the value of $\bpi_\e$ is qualitatively similar.

In order to assess which, if either, of the 3L1S and 2L2S solutions
is preferred, we must first be able to evaluate the implied physical
parameters for each.

\section{{Color-Magnitude Diagram and Blended Light}
\label{sec:cmd+blend}}

The first step toward this goal 
is to measure $\theta_*$ from the position of the source
on the color-magnitude diagram (CMD).  
This allows one to determine $\theta_\e=\theta_*/\rho$
and $\mu_\rel=\theta_\e/t_\e$, which are both important inputs into this
calculation.  In the present case, we also measure the astrometric position
and calibrated flux of the blended light, which will also provide
important constraints for interpreting the lens system.

\subsection{{Baseline  Object}
\label{sec:baseline}}
We first note that the various light-curve models described
above all obey $F_{\rm base}/f_s>40$, where $F_{\rm base}\equiv f_s + f_b$.
Therefore,
to an excellent approximation, the ``baseline object'' is essentially
the same as the blended light.

\begin{figure}
\plotone{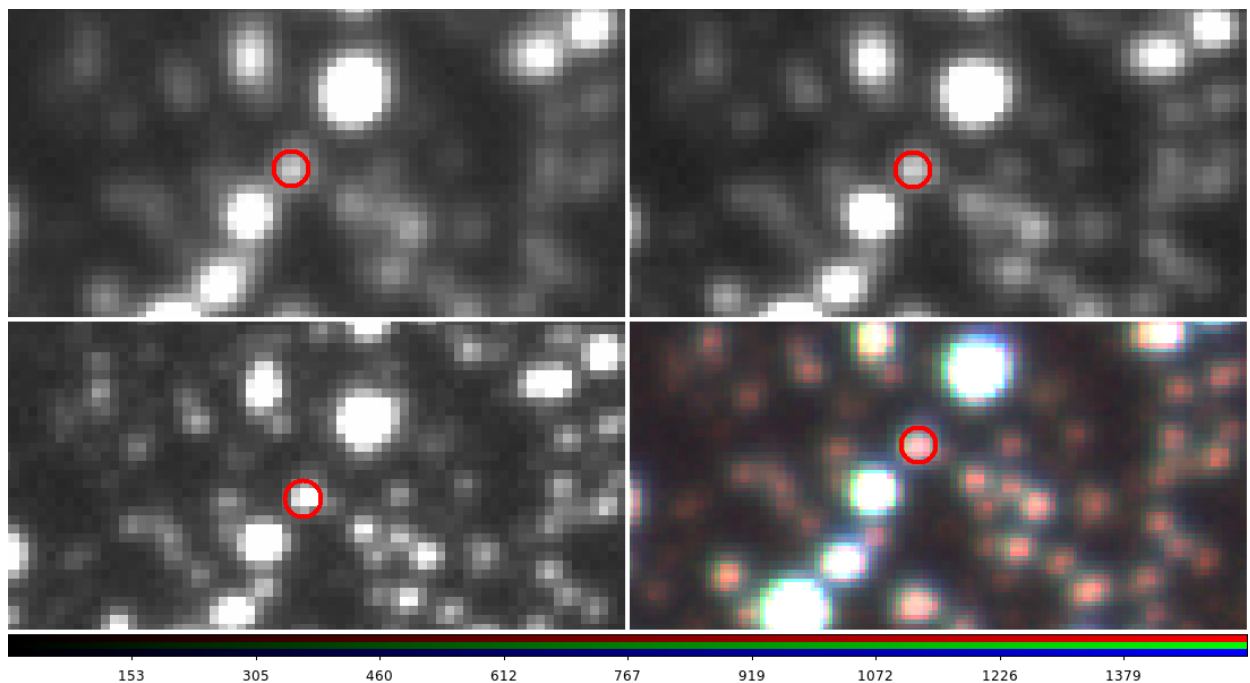}
\caption{CFHT images ($14^{\prime\prime}\times 6^{\prime\prime}$)
of the field near OGLE-2018-BLG-0532 in
$g$ (upper left), $r$ (upper right), $i$ (lower left),
and combined (lower right).  The red circle shows the
astrometric position of the microlensing event derived from
KMTC difference images near peak, transformed to the CFHT frame.
The ``baseline object'' shown here is nominally composed of the
source and blend, but the former contributes negligibly to the
total light.  The blend is $\la 50\,\mas$ from the lens, and
may be coincident, although this is less likely. The PSF of the ``baseline object'' (especially as seen in the $i$-band image) is not symmetric, but rather has a trefoil-like pattern to the North (up) and West (right) indicating two fainter stars are partially blended with the PSF.
See
Sections~\ref{sec:astrometry} and \ref{sec:blend}.
}
\label{fig:cfht}
\end{figure}

We conduct our investigation by analyzing relatively high-resolution
(FWHM $\simeq 0.45^{\prime\prime}$) images taken by the 3.6m Canada-France-Hawaii
Telescope (CFHT) from June to August in 2018.  
Because of higher resolution, these images might
in principle resolve out stars that are blended with the ``baseline object''
as seen in OGLE and/or KMT images.  Moreover, the higher resolution
and smaller pixel size permit a more precise astrometric measurement.
See Figure~\ref{fig:cfht}.

\subsubsection{{Astrometry}
\label{sec:astrometry}}

Here our goal is to find the astrometric offset between the microlensed
source and the baseline object.  If this is consistent with zero,
then the blend (essentially the same as the baseline object) could be
the lens.  And if this offset is small (and possibly zero),
it would be evidence that the blend was associated
with the event, i.e., either the lens itself, a companion to the
lens or a companion to the source.  On the other hand, if the baseline
object were well displaced from the microlensed source, then the
lens light could at most comprise a fraction of the baseline light, which 
would lead to stronger upper limits on the lens flux (see Appendix).

There are three steps to measuring this offset.  First, we must
find the position of the microlensed source relative to neighboring
field stars from difference images near peak.  Second, we must
find the position of the CFHT baseline object relative to these
same neighboring stars.  Third, we must transform the microlensing-template
image coordinates to the CFHT image coordinates.  We discuss each
of these procedures, with a focus on the error estimates of each.

In fact, we have carried out the entire procedure described below twice:
once using the source position on the KMT template and once using the
source position on the OGLE template.  For clarity, we report the 
KMT analysis first, and then the OGLE analysis.  Finally, we investigate
the origin of the differences.


{\bf Source Position:} As part of the normal process of image subtraction,
each image is astrometrically aligned to the template image, which
is then subtracted from it.  The difference image then basically
consists of an isolated star.  We measure this position for 12 highly
magnified KMTC images.  This sample has standard deviations in the
$(x,y)$ (essentially west and north) directions of $(0.0435,0.0320)$ pixels,
which are 400 mas.  Hence, the standard errors of the mean of this
measurement are $(5,4)\,$mas.  For completeness we note that the
correlation coefficient is 0.43, although this has no practical impact.

{\bf Baseline Object Position:}  In contrast to the microlens difference
image, which consists of a single star that is magnified and hence relatively
bright, the baseline image is filled with ambient stars.  Hence, in principle,
the measurement could be corrupted by ambient light either from 
gradients from relatively near stars, or from very nearby, semi-resolved 
stars.  Fortunately, we find that the baseline object is isolated from
relatively bright stars that could cause the first problem.  We do find
two semi-resolved very-nearby faint stars.  However, we ignore them here
and treat them further below under ``systematic errors''.

We construct transformations between pairs of CFHT images, one with
excellent (``best'') seeing and the other with good seeing.  We
measure the scatter in the positions of stars with similar magnitude
as the baseline object, finding 0.06 pixels, each pixel being 185 mas.
We attribute these as equally due to the two images, i.e., 0.04 pixels each.  
Using
the five best images (and conservatively using the scatter of these
measurements, rather than the slightly smaller mean scatter of similar
stars mentioned above), we derive standard errors of the mean 
$(6,4)\,$mas in the same $(x,y)$ directions.

{\bf Transformation:}  The transformation from the KMT to CFHT images
uses the {\sc Fitsh} {\sc grmatch} and {\sc grtrans} routines~\citep{Pal2012}
on stars within $1^\prime$
that are substantially brighter than the baseline object.
We find a scatter of these stars in the transformation of about 8 mas.
While roughly 200 stars were used to derive the transformation, we
conservatively consider only the 50 closest to the baseline object.
Then the error in the transformation is $\sim 8/\sqrt{50}\sim 1\,$mas.
This error is negligible compared to the other errors, and so we ignore it.
Hence, finally we derive
\begin{equation}
\Delta\theta_{\rm KMT}(E,N)=\btheta_{\rm base}-\btheta_{\rm source} = 
(-43,+20)\pm(6,4)\,\mas
\label{eqn:astroffset}
\end{equation}

{\bf Systematic Errors:}  Equation~(\ref{eqn:astroffset}) seems to
imply a highly significant $(\Delta\chi^2 \ga 50)$ difference in the
positions of the baseline object and source, indicating that the
baseline object is displaced to the west and north of the source.
However, one can see from Figure~\ref{fig:cfht} 
that there are two semi-resolved stars,
which are not registered by the DoPhot \citep{dophot,AlonsoGarcia2012}
astrometry of the CFHT image, that lie to the
west and north.  These are certainly biasing the tabulated position
of the source in these direction, particularly in the CFHT $i$ image.
However, it is not possible to evaluate the magnitude of these errors
due to the ``black-box'' nature of DoPhot astrometry.

{\bf Comparison to OGLE}:  We repeat the entire procedure beginning
with the OGLE-based source position and transforming this to the
CFHT image using the OGLE template and find
\begin{equation}
\Delta\theta_{\rm OGLE}(E,N)=(-61,+31)\pm(5,5)\,\mas .
\label{eqn:astroffset2}
\end{equation}
While the difference between Equations~(\ref{eqn:astroffset})
and (\ref{eqn:astroffset2}) is relatively modest relative to the errors,
it does have implications for the interpretation of the event.  Therefore
we investigate it further.

We first note that the OGLE template is from 2010 while the KMT template
is from 2018. Because the magnified source is only a factor 
$\eta\simeq 1.5$ times brighter than the baseline object, the apparent
position of the ``difference source'' is displaced by
$+\bmu_{\rm base}\Delta t/\eta$ relative to the true position of the source,
where $\bmu_{\rm base}$ is the proper motion of the baseline object relative
to the reference frame and $\Delta t=8\,{\rm yr}$.  Hence, to account for
the difference between the OGLE- and KMT-based offsets in the east direction
(the only one for which there is even a marginally significant discrepancy),
would require a proper motion of $\mu_{\rm base,E}=+3.4\pm 1.5\,\masyr$.
This is a very plausible number.  Hence, even the marginal discrepancy
between the OGLE and KMT values is entirely consistent with the KMT
value being correct.  Because the KMT reference image is from the same year as the event, the KMT-based procedure does not suffer
from the added uncertainty of the unknown proper motion of the baseline
object. Thus, we adopt Equation~(\ref{eqn:astroffset}).

We conclude that $|\Delta\btheta|< 50\,\mas$ with the balance of evidence indicating that the
baseline object is at least somewhat displaced from the lens. However, due to systematics, the displacement could be substantially
smaller than $50\, \mas$ and zero is not ruled out. Hence, the blend could be the lens, but more likely a large fraction of this light is due to another star.

\subsubsection{{Photometry}
\label{sec:photometry}}

We calibrate the CFHT instrumental $(g,i)$ photometry 
\citep{cfht-phot}
to calibrated 
OGLE-III $(V,I)$ photometry \citep{oiiicat}
to derive $I = i - 0.328 -0.054(g-i)$
and $(V-I)=0.326 + 0.716(g-i)$.   Combining these with the instrumental
magnitude measurements $(g,i)_{\rm base} = (19.992,18.596)\pm(0.014,0.023)$,
yields 
\begin{equation}
[(V-I),I]_{\rm base} = (1.362,18.596)\pm(0.019,0.024) .
\label{eqn:photometry}
\end{equation}

\subsection{{CMD}
\label{sec:cmd}}

We will ultimately wish to place the source color and magnitude
$(V-I,I)_s$ on an instrumental color-magnitude diagram in order
to find its offset from the centroid of the red clump and so
infer its angular radius $\theta_*$ \citep{ob03262}.  This in turn
will yield the angular Einstein radius via the relation
$\theta_\e=\theta_*/\rho$, where $\rho$ is the normalized source
radius, which is a parameter of the microlens modeling.

While $I_s$ can only be determined by conducting detailed
modeling of the microlensing event, $(V-I)_s$ is independent
of the model, and in fact can often be determined by the regression
of the $V$ on $I$ fluxes without any model at all\footnote{This is not
strictly true for 1L2S or 2L2S events in which the source stars have
different colors.  However, we will show that this has no significant
impact in the present case.}.  In the present
case, we will be considering models with somewhat different values of
$I_s$.  The purpose of this section is therefore to derive a
scaled relation for $\theta_*$ that is valid for all of these.
This is feasible because for fixed source color (and hence
fixed inferred source surface brightness), $\theta_*\propto f_s^{1/2}$.

Hence, we will simply adopt a fiducial source flux $f_{s,\rm fid}=0.020$,
which corresponds to $I_{s,\rm fid} = 22.25$ on an $I=18$ flux scale.
This value corresponds to the source flux in the ``standard'' 2L1S model
in Table~\ref{tab:2L1S}, but the choice of this particular value
is simply a matter of convenience.  Then, other than (temporarily)
treating the error in $I_s$ as zero, the analysis proceeds in the usual
way.

\begin{figure}
\plotone{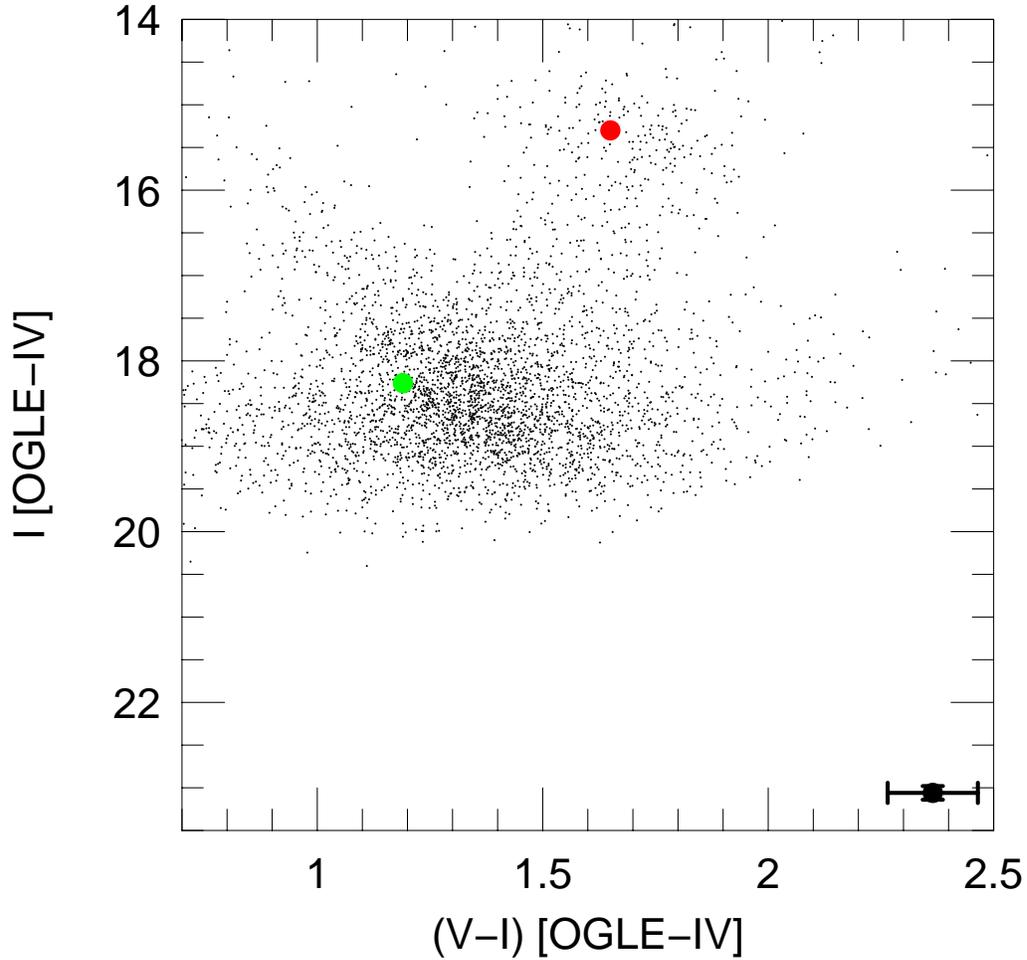}
\caption{OGLE-IV color-magnitude diagram (CMD) for OGLE-2018-BLG-0532.
The source star, blended light, and clump centroid are shown
as black, green, and red circles, respectively.  These results are
combined with a similar CMD derived from KMTC43 data to measure the
offset of the source star from the clump and so to derive the
angular source radius, $\theta_*$.  The astrometry and photometry 
of the blended light is discussed in Section~\ref{sec:baseline},
based on CFHT images shown in Figure~\ref{fig:cfht}.
}
\label{fig:cmd}
\end{figure}

We begin by constructing the OGLE-IV instrumental CMD 
(Figure \ref{fig:cmd}) and measuring
the position of the red clump centroid 
$[(V-I),I]_{cl,\rm OGLE-IV}=(1.65,15.30)\pm (0.04,0.08)$.  
Next we measure the source color 
$(V-I)_{s,\rm OGLE-IV}=2.36\pm 0.10$.  In practice we do this from
the models rather than regression, but we confirm that the measurement
and error is the same from the different models (as expected).  
Together with the adopted fiducial 
$I_{s,\rm fid}=22.25$, these imply an offset from the clump of
$\Delta [(V-I),I]_{\rm OGLE-IV}=(0.71,6.95)\pm (0.11,0.08)$.  Next,
we must take account of the fact that while OGLE-IV $I$ band is almost
exactly standard, OGLE-IV $(V-I)$ requires a correction factor of 0.93
to match the standard color\citep{ogle-iv}. Therefore, 
$\Delta [(V-I),I]_{\rm standard,OGLE}=(0.66,6.95)\pm (0.10,0.08)$.  

We repeat a similar procedure with KMTC43 data and obtain
$\Delta[(V-I),I]_{\rm standard,KMT}=(0.69,6.86)\pm(0.10,0.08)$.  These results
are consistent at the $1\,\sigma$ level.  Hence, we average them
and obtain $\Delta[(V-I),I]=(0.67,6.91)\pm(0.07,0.08)$.  

We adopt $[(V-I),I]_{0,{\rm clump}} = (1.06,14.39)$ from \citet{bensby13}
and \citet{nataf13}, which yields
$[(V-I),I]_{0,s} = (1.73,21.30)\pm (0.07,0.08)$, where we are ignoring
for the moment any errors in the intrinsic position of the clump.  
Then converting from
$V/I$ to $V/K$ using the color-color relations of \citet{bb88} and then
applying the ``dwarf and sub-giant''
color/surface-brightness relations of \citet{kervella04},
we obtain
\begin{equation}
\theta_{*,\rm fid} = 0.388\pm 0.019\,\muas.
\label{eqn:thetastar}
\end{equation}
As noted above, the error bar in Equation~(\ref{eqn:thetastar})
reflects only the errors in centroiding the clump and in measuring the
source color.  In particular, it does not include the measurement error
of $f_s$, nor does it include possible systematic errors
in the color/surface-brightness relations or the intrinsic position
of the clump.  To take account of all of these errors, we add 5\% in
quadrature to the $\simeq 5\%$ error reported in Equation~(\ref{eqn:thetastar}),
i.e., 7\% in total.
Because we discuss other solutions with somewhat different source fluxes
$f_s$, we note that these also have somewhat different $\theta_*$.  Including
this additional 5\%, the general formula for $\theta_*$ becomes
\begin{equation}
\theta_* = \sqrt{f_s\over 0.020}(0.388\pm 0.027\,\muas).
\qquad
\label{eqn:thetastargen}
\end{equation}
We note that the best-fit values for $\theta_\e=\theta_*/\rho$ for the
10 models considered in Tables~\ref{tab:2L1S}, \ref{tab:3L1S}, 
and \ref{tab:2L2S} are therefore 
$\theta_\e=(1.23, 1.35, 1.34)\,\mas$,
$\theta_\e=(1.30, 1.14, 1.20, 1.23)\,\mas$, and
$\theta_\e=(1.29, 1.43, 1.48)\,\mas$, respectively.

\section{{Physical Parameters}
\label{sec:physical}}

We now determine the physical parameters, such as the mass and distance
of the lens system,
\begin{equation}
M={\theta_\e\over \kappa\pi_\e},
\qquad
D_L={\au\over \theta_\e\pi_\e + \pi_s},
\label{eqn:massdist}
\end{equation}
for eight solutions presented 
in Tables~\ref{tab:2L1S}, \ref{tab:3L1S} and \ref{tab:2L2S}.  
We directly evaluate 
these from the output of the MCMC.  For example, the projected physical
separation of a given MCMC realization of the 2L1S solutions is
\begin{equation}
a_\perp\equiv s D_L\theta_\e = {s\cdot\au\over \pi_\e + \pi_s/\theta_\e}
= {s\cdot\au\over (\pi_{\e,N}^2+\pi_{\e,E}^2)^{1/2} 
+ \pi_s\rho/[\theta_{*,\rm fid}(f_s/0.020)^{1/2}(1+\delta)]}
\label{eqn:aperp}
\end{equation}
Because $\pi_\e\theta_\e=\pi_\rel\gg \pi_s$ (i.e., the lens is much closer
to the Sun than to the source), we treat the source distance as a 
constant $D_S =\au/\pi_s = 7.9\,\kpc$.  The ``$\delta$'' in the denominator
represents the 7\% error in $\theta_*$.  It is implemented by integrating
over a Gaussian for each MCMC realization.  The remaining variables
$(s,\pi_{\e,N},\pi_{\e,E},\rho,f_s)$ come directly from the MCMC.

Tables~\ref{tab:2L1Sphys}, \ref{tab:3L1Sphys} and \ref{tab:2L2Sphys} 
show the resulting
physical parameters for the 2L1S and 3L1S solutions, respectively.
To scale the projected separation to the snow line, we adopt
$a_{\rm snow} = 2.7(M/M_\odot)\au$.  The heliocentric and local standard of
rest (LSR) quantities are derived by
\begin{equation}
\bmu_\hel = \bmu_\geo + {\pi_\rel\over\au}\bv_{\perp,\oplus}:
\qquad
\tilde \bv_\hel \equiv {\au\over\pi_\e}\,{\bmu_\hel\over \mu_\hel^2}
\label{eqn:helio}
\end{equation}
and
\begin{equation}
\tilde \bv_{\rm lsr}  = \tilde \bv_\hel + \bv_{\perp,\odot}.
\label{eqn:helio2}
\end{equation}
Here $\bmu_\geo=\bmu_\rel$ is the geocentric relative proper motion,
$\bmu_\hel$ is the heliocentric proper motion, 
$\bv_{\perp,\oplus}(N,E)= (2.7,10.2)\kms$ is Earth's velocity projected onto
the plane of the sky at $t_0$, 
$\bv_{\perp,\odot}(l,b)= (12,7)\kms$ is the Sun's peculiar velocity projected onto
the plane of the sky, and $\tilde\bv_\hel$ and $\tilde\bv_{\rm lsr}$ are the
``projected velocities'' \citep{gould92} in the heliocentric and LSR frames,
respectively.

The physical properties of the host-Neptune system are quantitatively similar 
for the eight solutions shown in Tables~\ref{tab:2L1Sphys},
\ref{tab:3L1Sphys}, and \ref{tab:2L2Sphys}
except for the direction of $\bmu_\hel$.  This follows from the similarity
of the underlying microlensing parameters in 
Tables~\ref{tab:2L1S}, \ref{tab:3L1S} and \ref{tab:2L2S}.  
It is expected that the
four pairs of $[(u_0<0),(u_0>0)]$ will be similar to each other
(apart from the direction of $\pi_\e$).  The fact that the properties
of the Neptune do not change much when the light curve is fit for
an additional planet conforms to the prediction of \citet{zhu-2planet},
who found from simulations that failure to take account of 
``unrecognized'' second planets does not generally have major impact
on the parameter estimates of the first planet.  However, \citet{zhu-2planet}
did not simultaneously fit for parallax, so the fact that the estimate
of $\bpi_\e$ is robust against the presence of a planet with a 
$\Delta\chi^2\sim 80$ signature could not necessarily have been anticipated.

The host mass is estimated at 0.20--0.25 $M_\odot$, at a distance 
$D_L\sim 1\,\kpc$.  At this mass, its absolute magnitude should be $M_I\ga 9$,
and so allowing for an extinction of $A_{I,l}\sim 0.5$ mag to this distance, 
the predicted lens flux is $I_l\ga 19.5$.  This is below the limit
from the blended light from Equation~(\ref{eqn:photometry}), $I_l>18.6$.
Hence, the blended light appears to be dominated by some other star,
which would be consistent with the ``balance of evidence'' in
Section~\ref{sec:astrometry} that the baseline object is offset from the
lens.  Nevertheless, the fact that the predicted lens flux is of the
same order as the blend flux does raise some subtle issues, the discussion
of which we defer to Section~\ref{sec:discuss}.

The first (robustly detected) planet has a mass of 6--10 $M_\oplus$ across
the six solutions, which is near the mass $M_p\sim 10\,M_\oplus$ generally
thought to be required for rapid growth by gas accretion.  We also
leave the implications of this mass estimate to the discussion.  It lies
projected at about twice the snow line distance.

The second (putative) planet has a mass midway between Jupiter and Saturn.
It lies projected either at $\sim 0.8$ or $\sim 5.6$ times the snow
line, depending on whether the close or wide solution is correct.
The close solution is very slightly preferred by $\chi^2$.  From
general statistics of such gas giants, they are more likely to be
found outside than inside the snow line, but for this individual
case we cannot distinguish.

\section{{Discussion}
\label{sec:discuss}}

The OGLE-2018-BLG-0532L system is interesting for two major reasons.
First, it shows significant evidence for a second planet, which would make
it only the third two-planet system discovered by microlensing.
Second, its securely detected planet, OGLE-2018-BLG-0532Lb
lies near the possible ``pile-up'' of planets identified by \citet{kb170165}
that is centered just below $q=10^{-4}$.  In addition, the blended light
in this event raises some puzzles that deserve further discussion.

\subsection{{Two-Planet System?}
\label{sec:twoplan}}

While the ``first planet'' (the Neptune) in this system
is readily apparent from the light curve, even without detailed
modeling, visual inspection of the light curve does not
give even the slightest hint of the ``second planet'' (the Jupiter).
This contrasts sharply with the first two cases of microlensed
two-planet systems, OGLE-2006-BLG-109Lb,c \citep{ob06109,ob06109b}
and OGLE-2012-BLG-0026Lb,c \citep{ob120026}, wherein the light curves 
basically ``factored'' into perturbations due to each planet separately
\citep{planet-factor}.
The present case is more similar to the simulations of two-planet systems 
in KMTNet-like data streams that were investigated by \citet{zhu-2planet},
in which the ``second-planet'' signature was often a weak featureless
deviation.  Nevertheless, while these second-planets were often not obvious
to the eye, they did clearly stand out in residuals to the fits to the
first planet.  This is not true in the present case.

Rather, as discussed in 
Appendix~\ref{sec:append3L1S}, 3L1S solutions
were only investigated because some intermediate results of the
analysis seemed inconsistent with the upper limits on lens flux.
This investigation consumed a lot of resources (both human and
computer), so such investigations 
are not generally undertaken in the analysis of planetary
microlensing events in the absence of discernible evidence for a third
body.  As we noted in Section~\ref{sec:intro}, however, \citet{shin-3L1S}
did search eight archival planetary events for evidence of third bodies
and found $\Delta\chi^2\ga 50$--142 in three of these cases, which are
similar to the $\Delta\chi^2=81$ improvement for OGLE-2018-BLG-0532.

We argue that this planetary signal is plausibly real, but we consider
that the evidence for the planet is not sufficient to definitively
claim its detection.  In this regard, there are two distinct
questions that must be examined.  First, is the $\Delta\chi^2=81$
difference between 3L1S and 2L1S secure enough so that this improvement
cannot plausibly attributed to systematics?  Second, is the $\Delta\chi^2=16$
preference of 3L1S over 2L2S high enough to confidently claim that the
additional ``body'' relative to 2L1S is an additional lens (3L1S) rather
than an additional source (2L2S)?  We will answer ``yes'' to the first
question and ``no'' to the second.

Regarding the preference of 3L1S to 2L1S, there are three factors
that all favor accepting the 3L1S solution.
First, the $\Delta\chi^2=81$ improvement is relatively high.
It is true that various authors have advocated much higher thresholds
for ``detectability'' of planets, but the reasons for this higher
threshold must be clearly understood.  For example, \citet{gould10}
advocated a range of possible thresholds centered on $\Delta\chi^2=500$.
However, their argument was not that this was the minimum required
to detect, or even to characterize a planet.  Rather, it was the minimum
$\Delta\chi^2$ using {\it re-reduced data} needed
to guarantee that a planet would be sufficiently apparent in
{\it pipeline data} to trigger a deeper investigation.  Without
this ``guarantee'' that all planets meeting some objective criterion
have been found in a given sample, one cannot carry out a statistical 
analysis of the sample.  But the problem of finding a second planet
in a system already known to contain one is quite different from 
that of finding the first planet.  First, the data are {\it already}
re-reduced.  Second, the sample (events with planets) that is being
probed for second planets is much smaller than the sample (all events)
that was probed for first planets.  Hence, the probability of
some unknown systematics introducing a ``planetary signal'' is likewise
reduced.  Finally, in the present case, we are not trying to
create a statistical sample, but only to assess the evidence that
the planet is real.  Therefore there is no need to set a threshold
that is ``high enough'' that the great majority of planets meeting
it would be recognized ``by eye''.

\begin{figure}
\plotone{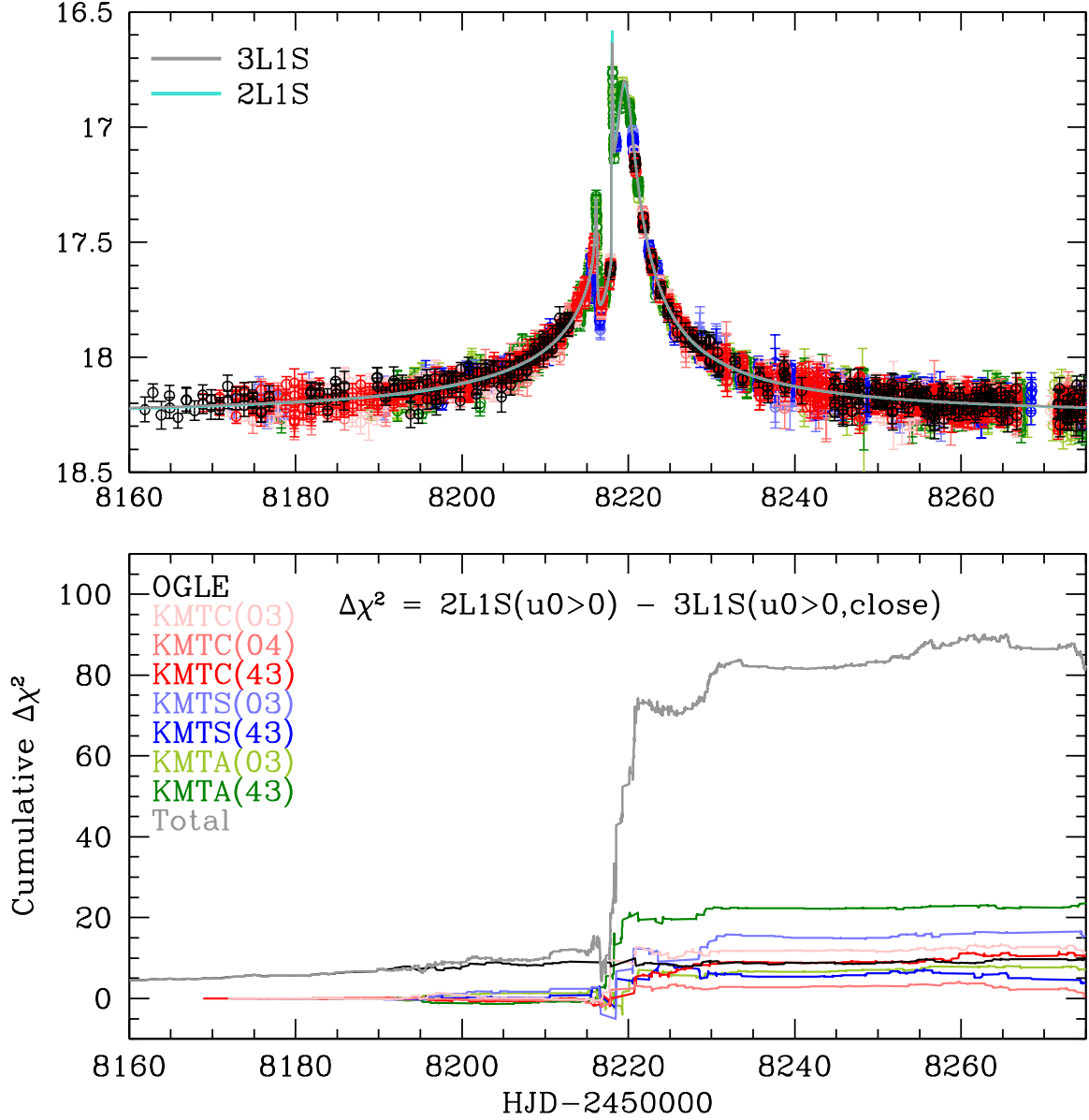}
\caption{Cumulative $\Delta\chi^2(t) = 
\chi^2_{{\rm 2L1S},(u_0>0)} (t) - \chi^2_{{\rm 3L1S},(u_0>0,{\rm close}} (t)$ 
function for the addition of a second planet in OGLE-2018-BLG-0532.
All eight observatory/field combinations contribute positively, and most
of the signal comes from the region of the anomaly.  Both of these facts
contribute to the confidence that the putative second planet is not due to systematics (however, see Section \ref{sec:2L2S}).
}
\label{fig:dchi2}
\end{figure}

\begin{figure}
\plotone{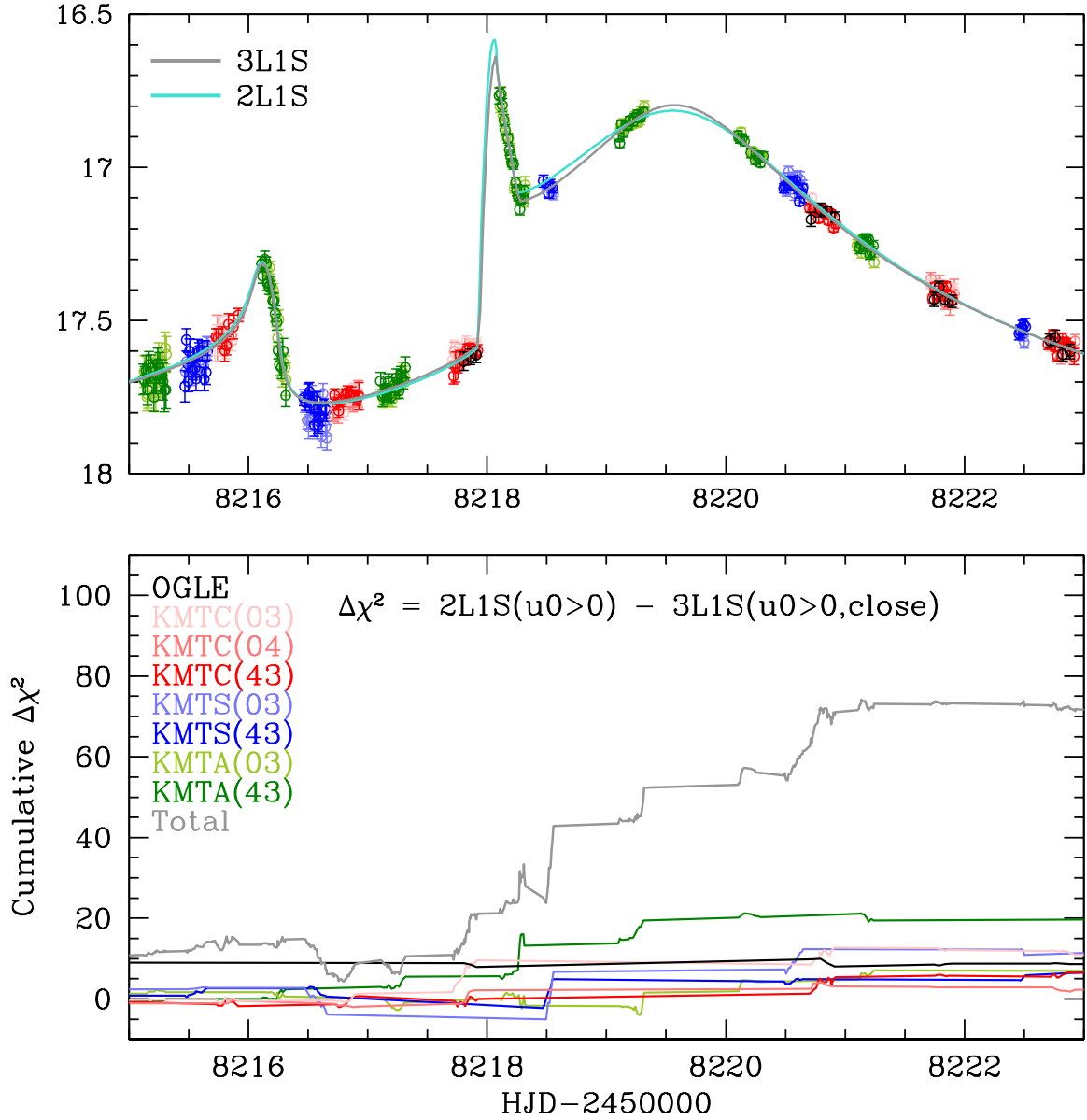}
\caption{Zoom of the cumulative 
$\Delta\chi^2(t) = \chi^2_{\rm 2L1S}(t) - \chi^2_{\rm 3L1S}(t)$ 
function shown in Figure~\ref{fig:dchi2}.}
\label{fig:dchi2part}
\end{figure}

Second, the cumulative distribution 
$\Delta\chi^2(t) = \chi^2_{\rm 2L1S}(t) - \chi^2_{\rm 3L1S}(t)$ 
is quite consistent with a real planet and would require a remarkable
set of coincidences if it were due to systematics.  As shown
in Figure~\ref{fig:dchi2}, all eight observatory/field combinations
contribute positively to the total $\Delta\chi^2$. Moreover, as
would be expected, most of the contribution is from the regions
of the caustic structure, where a weak shear (or pseudo-shear) would
generate the most pronounced effects.
See Figure~\ref{fig:dchi2part}.
And again, within this critical region, all observatory/field combinations
contribute positively.  Moreover, the whole light curve
does weakly contribute as well, also as one would expect.\footnote{In principle, all eight
observatory/field combinations could also contribute positively if
there were a common physical cause due to unmodeled physics of the
microlensing event or coherent variability of the source or blended light.
As we discussed in Section~\ref{sec:2L1S} and \ref{sec:3L1S}, orbital
motion does not give rise to significant signals during the event.  The
blend is an upper main-sequence or turnoff star and so is not expected
to be significantly variable.  Nevertheless, we search for such putative
variability in binned residuals to the 2L1S fit, but find no coherent
deviations except near the peak.  Source variability at the required
level could not be detected except when highly magnified
(i.e., near the peak) because it is too faint.  However, the source is
a late K or early M dwarf and so is also not expected to vary on few
day timescales.  Moreover, a minority (but still significant) part of
the 3L1S signature comes from the long-term behavior of the event, which
would require that the source vary on two different timescales in a
``cooperating'' fashion.  We conclude that all such explanations by
real physical effects are unlikely.}.    

Third, a weak signal from a second planet in a high magnification
event should not be regarded as unexpected. As discussed in 
Section~\ref{sec:intro}, high-magnification events are simultaneously
sensitive to all planets in the system, provided that they are close
enough to the Einstein ring to generate a sizable central caustic.
At the time of the
\citet{shin-3L1S} study,
there were a total of nine ``high-magnification'' ($u_0<0.01$)
events that contained published planets.  This set consists of six
of the eight events analyzed by \citet{shin-3L1S} (i.e., not including
OGLE-2005-BLG-071 and MOA-2009-BLG-387, which had $u_0>0.01$), plus three additional events:
OGLE-2006-BLG-109 \citep{ob06109,ob06109b},
OGLE-2012-BLG-0026 \citep{ob120026}, and
OGLE-2007-BLG-349 \citep{gould10,ob07349}.
The first two of these three additional events showed 
very clear evidence of two planets, while the last had very clear
residuals from the 2L1S fit due to a binary companion to the host,
i.e., circumbinary planet.  Of the six events that did not
have discernible systematic residuals, i.e., did not
require a third body to achieve a satisfactory fit to the light curve,
\citet{shin-3L1S} found that two had evidence for a third body
at a similar or higher $\Delta\chi^2$ as OGLE-2018-BLG-0532.
Further, one other had evidence at a lower level ($\Delta\chi^2=50$)
that still could be a plausible candidate.  While \citet{shin-3L1S}
did not investigate all of these marginal detections in detail,
they did mention that three different observatories contributed
significantly to the $\Delta\chi^2=142$ of their best case.  Stated
otherwise, only three of the nine cases in this complete sample of
high-magnification planetary events showed no significant evidence 
$(\Delta\chi^2<30)$ for third bodies.

To further illuminate this issue, we carry out simulations based on
OGLE-2018-BLG-0532 to determine how often one would expect
``weak-but-detectable'' signals versus ``strong-obvious'' signatures,
assuming that a Jovian-mass-ratio planet like the possible second planet in OGLE-2018-BLG-0532
was somewhere in a system basically defined by the robustly-detected
planet OGLE-2018-BLG-0532Lb.  To do so, we create simulated
light curves following the procedure of \citet{ob171434}.  That
is, we measure the residuals from the 3L1S model and add these
to models with the same $q_2$ but different $s_2$ and $\psi$ compared to the
best fit model.  The seven standard parameters 
$(t_0,u_0,t_\e,s_1,q_1,\alpha,\rho)$ are kept the same.  Then we fit to
a standard 2L1S model (i.e., with seven parameters) and find the
increase in $\Delta\chi^2$ due to the absence of the second planet
in the model.  Note that to enhance the computational speed, we
both create 3L1S models and fit to 2L1S models with $\bpi_\e=0$.
Conceivably, this might alter the $\Delta\chi^2$ due to the additional
degrees of freedom.  However, this seems unlikely on general grounds
and in the few cases that we checked, the fractional change in
$\Delta\chi^2$ was a few percent, i.e., well below the level
of interest for this theoretical study.

\begin{figure}
\plotone{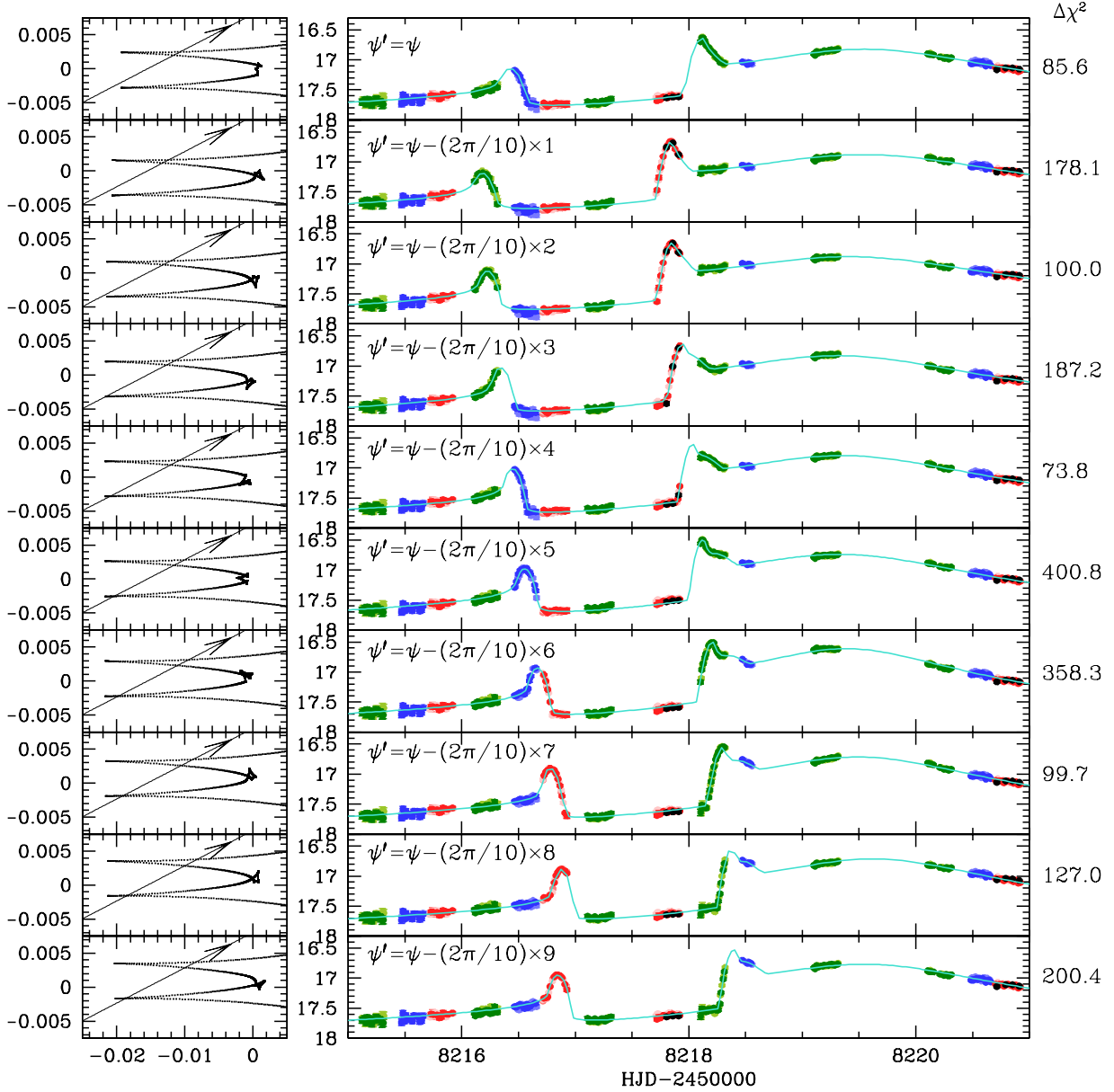}
\caption{Simulated 3L1S light-curve data (colored points) with best-fit 2L1S
models (cyan curves), for separation $\ln s_2=1$ and various values of
the angle $\psi^\prime = \psi -2\pi n/10$ ($n=0,1,\dots 9$) 
between the two planets, but with the second planet
having the same mass ratio as in the best fit.    The 3L1S 
caustic geometries are shown at left, and the 
$\Delta\chi^2 = \chi^2({\rm 2L1S}) - \chi^2({\rm 3L1S})$ values are shown
at right.  The $n=5$ and $n=6$ cases show noticeable offsets of the
data from the models, but in other cases the differences are difficult
to discern by eye.}
\label{fig:fake_44}
\end{figure}

\begin{figure}
\plotone{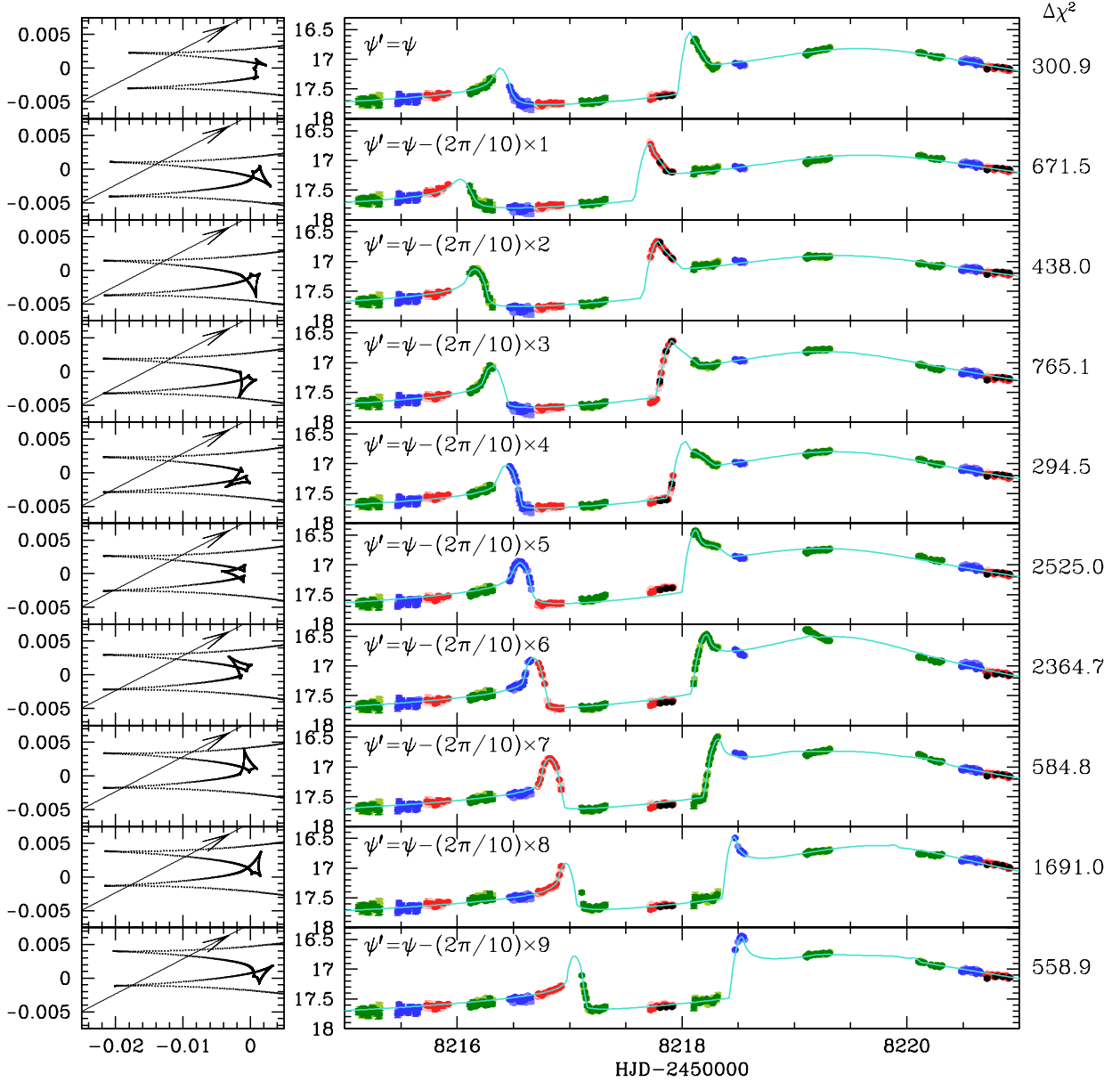}
\caption{Simulated 3L1S light-curve data with 2L1S models, similar to
Figure~\ref{fig:fake_44}, except that $\ln s_2=0.75$.  For $n=5$ and
$n=6$, the light-curve deviations strongly suggest a third body, and
several other panels show obvious deviations.
}
\label{fig:fake_34}
\end{figure}

\begin{figure}
\plotone{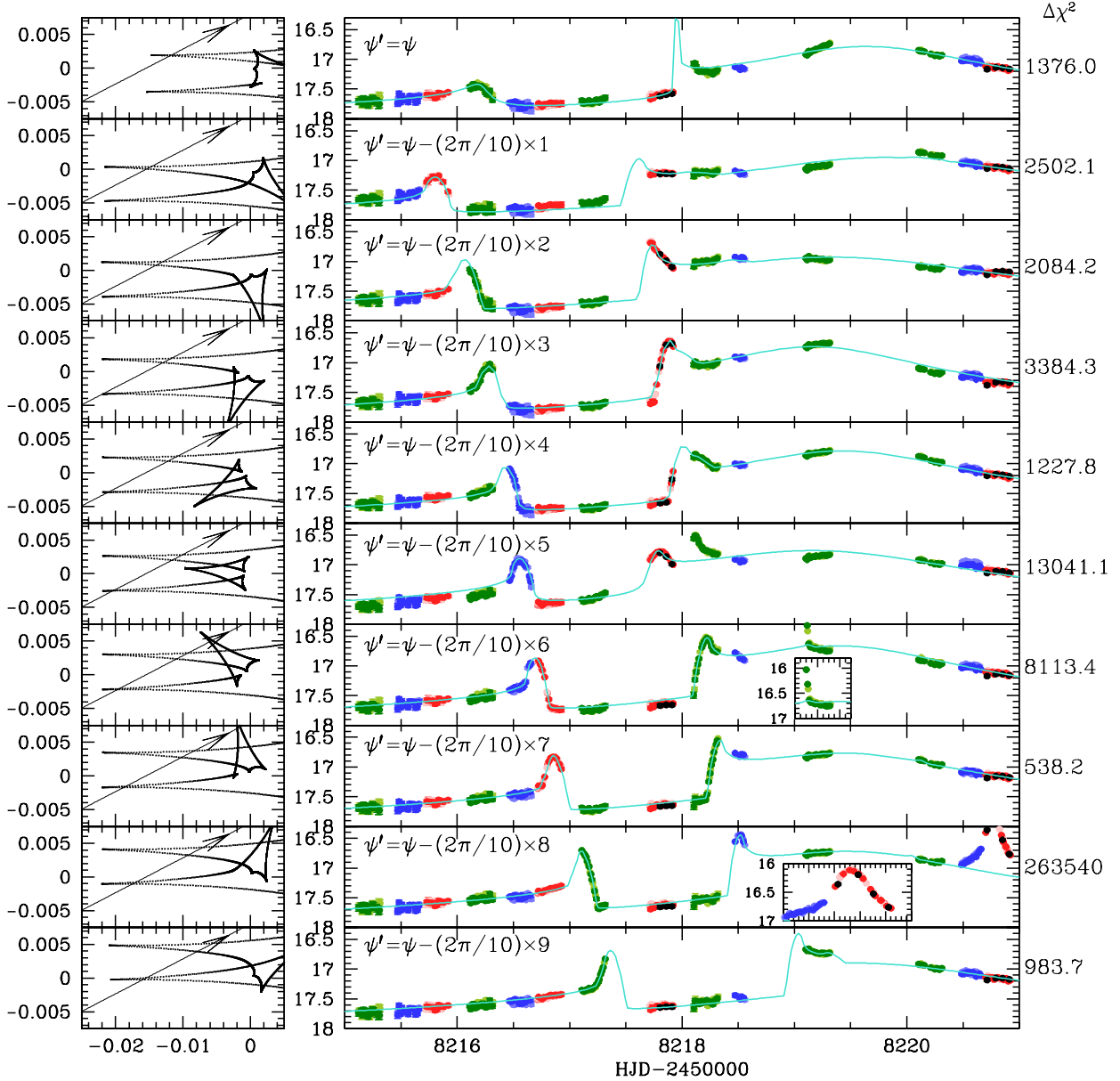}
\caption{Simulated 3L1S light-curve data with 2L1S models, similar to
Figure~\ref{fig:fake_44}, except that $\ln s_2=0.5$. For $n=6$ and $n=8$,
small insets show the excursions of the data beyond the standardized
window-size of the panel.  The cases $n=5$, $n=6$, and $n=8$ suggest that
the light curve ``factors'' into contributions from two planets, as described
by \citet{planet-factor}.
}
\label{fig:fake_24}
\end{figure}

Figures~\ref{fig:fake_44}, \ref{fig:fake_34}, and \ref{fig:fake_24}
show the results of these simulations for $\ln s_2 = (1,0.75,0.50)$,
i.e., $s=(e^1,e^{3/4},e^{1/2})$, respectively.  Note that the first
of these is very similar to the best fit value for the real data,
$s_2=2.65$.  In each case, we examine 10 values of 
$\psi^\prime = \psi_{\rm best} - 2\pi n/10$ with $n=0,1,\dots 9$.  For
each simulation we show the caustic geometry at the left and 
$\Delta\chi^2$ at the right.  We see that at $\ln s=1$, only
two of the 2L1S fits look ``clearly suspicious'' i.e., $n=5$ and
$n=6$.  The first shows strong systematic residuals in
KMTA and OGLE/KMTC data on 8217.xx.  The second shows such
residuals for OGLE/KMTC data on 8217.xx and KMTA data on 8218.xx.
However, none of the 10 cases show residuals that are obviously
due to a third body.

By contrast, for $\ln s_2=0.75$, there are two cases (again
$n=5$ and $n=6$) with residuals that clearly look like perturbations
due to a third body, although if one were sufficiently suspicious of the 
data one might be inclined to dismiss them at ``just systematics in 
the data''.  Only at $\ln s_2=0.5$ are there several cases
($n=5$, $n=6$, and $n=8$) for which the light curve appears
to be a superposition of two planetary perturbations, as
predicted by \citet{planet-factor} and as actually observed
in the cases of OGLE-2006-BLG-109 \citep{ob06109,ob06109b} and
OGLE-2012-BLG-0026 \citep{ob120026}.  However, for this $\ln s_2=0.5$
case, nearly all the examples have very obvious residuals that would
almost certainly prompt investigation for an additional body
or bodies in the system.  We expect that for second planets even
closer to the Einstein ring, $0<\ln s_2 < 0.5$, essentially all cases
would yield obvious residuals that would have to be investigated.  Further,
due to the $s\leftrightarrow s^{-1}$ degeneracy, we expect that
detectability would be symmetric with respect to the Einstein ring.

If we then consider potential planets that are uniformly distributed
in $\ln s$ (i.e., Opik's law), the instances of ``obvious signatures''
and ``significant but non-obvious'' signatures are roughly
comparable for the geometry and light-curve coverage of OGLE-2018-BLG-0532.
This study is meant to be only illustrative.  A full investigation
of the statistics of such weak signals would require systematic modeling
of all high-magnification events with planets.  This would be well
beyond the scope of the present paper.

 Thus, the answer to our first question is that based solely
      on comparison of the 3L1S to the 2L1S solutions, we would
      conclude that the apparent second (i.e., Jovian) planetary signal is not due to
      systematics.  
Given that the signature of this planet
was invisible to the eye, even in the residuals, we think that it
would be prudent to systematically search for third bodies in all
high-magnification planetary events.

Moreover, such a systematic search could reveal additional binary
companions to planetary systems.  To date, there are four microlensing
planets in microlens-binary systems, 
OGLE-2007-BLG-349 \citep{ob07349}, 
OGLE-2013-BLG-0341 \citep{ob130341},
OGLE-2016-BLG-0613 \citep{ob160613}, and  
OGLE-2008-BLG-092 \citep{ob08092}.  
The first two were in 
high magnification events.  For both of the
middle two, the binary companion provides by far the dominant
signal in the light curve, while for the first it generates very
noticeable residuals to the single-planet fit.  The last 
case has a completely different geometry, in which each of the
three bodies gives rise to a nearly isolated microlensing event.
Detection of additional systems
would be of interest in their own right.  In addition, inclusion
of the third (binary-companion)
body in the fit could change the parameters of the planet.

We turn now to the second question: can we confidently accept 3L1S
over 2L2S?  The main argument in favor is that the $\Delta\chi^2=16$
would virtually rule out ($p=3\times 10^{-4}$) 2L2S if the noise 
in the data were uncorrelated (``white'' rather than ``red'').
However, for microlensing data, it is well known that red noise exists
at some level.  Moreover, in contrast to the analogous case of
comparing 2L1S versus 1L2S, the complexity of the underlying model
would allow it to adapt to systematics in the data in ways that
are difficult to understand intuitively.  
Hence, in our view, a $\Delta\chi^2=16$
preference in this case would require a supplementary argument
for reliable confirmation.

Such a confirmation/rejection could come from a color test of the 2L2S
solution.  For example, \citet{ob151459} showed that the color-offset
between two of the three sources in the 1L3S solution of OGLE-2015-BLG-1459
clearly confirmed it over the otherwise degenerate 2L2S and 3L1S solutions.
In the present case, the two sources of the 2L2S solution
are both on the main sequence with the secondary being $\Delta I\simeq 0.8$
mag fainter than the primary.  Hence, one would expect that it would also be
redder by $\Delta(V-I)\simeq 0.35$.  On the other hand, if the 2L2S
solution were simply mimicking a lightcurve generated a 3L1S (i.e., 
single-source, hence achromatic) event, then we would expect
$\Delta(V-I)\rightarrow 0$.  Thus, this provides a potentially
clear test.

We therefore include all $V$ data into the fit and model these by
an additional parameter $q_{F,V}$ and evaluate
$\Delta(V-I) = 2.5\log(q_{F,I}/q_{F,V})$.  Unfortunately, due to the
redness of the sources as well as the relatively bright blue blend
and the relative paucity of $V$-band data (1/10 of $I$-band), 
this test does not
yield decisive results.  We find that the best fit is indeed
close to the 2L2S expected value of $\Delta(V-I)\sim 0.35$ and that 
3L1S value (zero) is disfavored by $\Delta\chi^2=3.5$ (for 1 dof).
Of course, such a small value $\Delta\chi^2$
cannot be regarded as strong evidence,
but it is a completely independent test that works in the direction
of undermining confidence of the 3L1S solution.

Considering all the evidence, we conclude that while the
second planet is plausibly real, it cannot be definitively claimed.

\subsection{{Cold Neptune}
\label{sec:neptune}}

OGLE-2018-BLG-0532Lb is the second cold Neptune to be discovered since the
recent analysis by \citet{kb170165} of the mass-ratio function
below $q<3\times 10^{-4}$.  They had argued for a sharp break in 
the mass-ratio function at $q_{\rm brk}\sim 0.56\times 10^{-4}$
and/or a ``pile up'' of cold Neptunes just above the putative
break point.  Both of these planets (the other being OGLE-2015-BLG-1670Lb,
\citealt{ob151670}),
have mass-ratio estimates whose error bars straddle $q=1.0\times 10^{-4}$.
This places them near the upper end of the putative ``pile-up''.
While it would be premature to revisit the form of the mass-ratio function
at this point, we note that the pace of discovery of planets in the
regime $q\la 1\times 10^{-4}$ is accelerating.  The discovery years
for the nine planets in this regime are (2005, 2005, 2007, 2009, 2013, 2015,
2016, 2017, 2018).  That is, the discovery pace has roughly doubled
starting in 2015.  This can at least be partly attributed to the
inauguration of KMTNet in that year in the sense that KMTNet played
a critical role in two of the four discoveries since 2015, and
an auxiliary role in the other two.  This tends to confirm that the
apparent doubling of the discovery rate for low-$q$ planets is real and
not the result of a statistical fluctuation.  Hence, it is likely
that within a few years we will gain a significantly better picture
of the mass-ratio function at the low end.

\subsection{{Nature of the Blended Light}
\label{sec:blend}}

The blended light in this event represents something of a puzzle.
The four facts that need to be evaluated are as follows. (1)  The source
contributes only a tiny fraction of the light from the ``baseline
object'', meaning that the ``baseline object'' can be be effectively
identified with the blend. (2) The blend is astrometrically offset
from the lens by $\la 50\,\mas$, but we concluded that the balance
of evidence was that it was not coincident with the lens.  (3) The
predicted flux from the lens in the best-fit models is about one mag
fainter than the blend, so the upper limits on lens light are clearly
satisfied.  (4) The blend is ``relatively  blue'' in a sense that
we will make clear shortly.

The blend must be either (1) the lens itself, (2) a companion to the
lens; (3) a companion to the source; (4) an unrelated ambient star; or
(5) combined light from two (or more) of the above, for which both
light sources contribute significantly. 

First, consider the scenario that the blend is behind the same dust column as 
the clump.
Recall from Section~\ref{sec:photometry} that 
$[(V-I),I]_{\rm base} = (1.36,18.60)$.
The extinction toward the clump is $[E(V-I),A_I]_{\rm cl} = (0.80,0.97)$
\citep{nataf13}.  
If the blend is in the bulge or more than a few kpc from us in the
disk, then it will suffer the same extinction. In this case,
$[(V-I),I]_{0,\rm base} = (0.56,17.63)$.  In particular, if the blend is
in the bulge, then $[(V-I)_0,M_I]_{\rm base} \simeq (0.56,3.1)$.  This
is quite consistent with a metal-poor turnoff star, particularly if
we take account of the roughly 0.05 mag combined color error from
the blend measurement and the estimate of $E(V-I)$.  Hence, considering
the blend light in itself, by far the simplest explanation is
that it is either a companion to the source or an ambient star in the
bulge.  The small astrometric offset between the blend and the
lensing event makes the first possibility more likely, but the two possibilities
are essentially the same in their implications for the lens system.

The problem with this hypothesis, however, is that the event models predict
that the lens itself contributes a significant amount of light
to the blend and that this light is red.  This means that the
``remaining light'' from the blend must be even bluer than just deduced.
That is, at $M=0.25\,M_\odot$ and $D_L\simeq 1\,\kpc$, the lens would have
$((V-I)_0,I_0)_L\sim (3.0,19.0)$.  Assuming, for example,
$(E(V-I),A_I)_L\sim (0.4,0.5)$ (i.e., half the dust column in front of the
lens), this implies $((V-I),I)_L\sim (3.4,19.5)$.  That is, the
lens would contribute about 44\% of the $I$-band flux, but only
7\% of the $V$-band light.  Hence, $((V-I),I)_{\rm remain}\sim (0.8,19.2)$.
Thus, if this remaining light were at the distance to (i.e., extinction of) the lens, it would
have $[(V-I)_0,M_I]_{\rm remain}\sim (0.4,8.7)$, while
if it were at the distance to (i.e., the extinction of) the source, it would have 
$[(V-I)_0,M_I]_{\rm remain}\sim (0,3.7)$. Neither of these positions on the CMD 
correspond to any normal star.  Nor would the problem be resolved
by placing the ``remaining light'' somewhere else along the line of
sight between the lens and the source.  Of course, we have made these 
calculations using one particular estimate for the lens mass and hence
$V/I$ fluxes.  However, if the lens is contributing any substantial
amount of light as a low-mass star near $D_L\sim 1\,\kpc$, then the
``remaining light'' will be very blue and difficult to explain.
This presents challenges for scenarios in which the blend is a companion to either the source or the lens or is an ambient star. 

Another possibility is that the blend is actually the lens.  This would
require two ``adjustments'' to the results of the measurements
described in the body of this paper.  First, the blend would
have to be coincident with the lens rather than being displaced
by $\la 50\,\mas$.  As we have discussed in some detail in
Section~\ref{sec:astrometry}, this is possible, although against
the ``balance of evidence''.  Second, the lens would have to be
substantially farther from us than the microlensing models predict.  
This is because the blue light
of the blend (if attributed to a single star) would require that
the star be much more massive than in the best solution, so much
brighter, and therefore much farther in order to be consistent
with the measured light from the blend.

Changing the mass and distance creates tension with the microlensing models because it requires changing either $\theta_\e$
or $\pi_\e$.  The more plausible route is to decrease $\pi_\e$ because
(from Equation~(\ref{eqn:massdist})) this both increases the mass and 
increases the distance as required by our hypothesis above.  In addition, $\theta_\e$ is much more robustly
measured than $\pi_\e$.  In particular, the amplitude $\pi_\e \equiv |\bpi_\e|$  is dominated by
$\pi_{\e,N}$, which has relatively large formal errors that are a direct
result of the fact that it is more difficult to measure than $\pi_{\e,E}$.
One quickly finds that the lens must lie at least
several kpc from us, so behind most or all of the dust.  This implies
$M\ga 1\,M_\odot$ to explain the color $(V-I)_{0,\rm base}\simeq 0.56$.  Then
$\pi_\rel = \theta_\e^2/\kappa M \rightarrow 0.135\,\mas (M/M_\odot)^{-1}$,
where we have adopted $\theta_\e = 1.05\,\mas$ for reasons that we discuss
below.  Thus, for the case $M=1\,M_\odot$, $D_L= 3.8\,\kpc$, and hence
the measured blend flux would imply $M_I =4.7$.  This is somewhat
dim for even a metal-poor turnoff star, but perhaps compatible
given the uncertainties.  The main remaining issue is that
this lens mass requires $\pi_\e=\theta_\e/\kappa M \rightarrow 0.13$
compared to $\pi_\e=0.47\pm 0.16$ for the lowest of the four parallax
solutions in Table~\ref{tab:3L1S}.  However, given the relatively
large errors, this ``large'' deviation may not be very seriously
disfavored.  Naively, it appears to be only a $2.1\,\sigma$ discrepancy.
In order to test this possibility more rigorously,
we force the ``close'' and ``wide'', $(u_0<0)$ geometries 
by fixing $\pi_{\e,N}=0.12$.
We find, indeed, that these solutions (Table~\ref{tab:forced}) are 
only disfavored relative to the corresponding solutions in Table~\ref{tab:3L1S}
by $\Delta\chi^2\sim 3$, and therefore by only $\Delta\chi^2\sim 7$ relative
to the best solution.
Given that there are often systematic errors in microlensing at this
level, these values do not in themselves disqualify this blend=lens
solution.  If future high-resolution imaging (see Section~\ref{sec:future})
shows that the blend
is closely aligned with the lens, then this solution will become the
most probable.  If not, it will be ruled out.

Finally, another possibility is that the parallax has been underestimated
by the fit, again due to low-level systematics.  For
example, if the lens were 1.5 mag fainter than we have estimated from
the best fit microlensing model (i.e., using Equation~(\ref{eqn:massdist})), then the lens would contribute only 11\% to the $I$-band
light, giving the ``remaining light'' a color of $(V-I)_{0,\rm remain}\sim 0.46$.
This is still quite blue, but allowing for measurement errors, marginally
acceptable, which would resolve the problems with supposing the blend is an unrelated star or a companion to the lens or source.  
The change in $\pi_{\e,N}$ required to produce a such a fainter lens would induce similar modest
stress on the fit as the one proposed above.  It is intrinsically 
less likely because there are fewer nearby stars in the observation
cone than distant stars.  However, if future high-resolution images
show that the lens is displaced from the blend, then it will become
more probable.

\subsection{{Future High Resolution Imaging and Spectroscopy}
\label{sec:future}}

Many puzzles remain for the OGLE-2018-BLG-0532L system, at least some
of which can be clarified by high resolution followup observations,
either adaptive optics (AO) observations from the ground or possibly
{\it Hubble Space Telescope (HST)} or {\it James Webb Space Telescope (JWST)}
observations from space.

The foremost question is the relation of the blended light to the event.
Because the blended light is projected $\Delta\theta\la 50\,\mas$
from the lens, it is very likely to be associated with the event, i.e.,
either the lens itself, a companion to the lens, or a companion to the
source.  However, in principle, the blend could be an unrelated ambient
star.

The most exciting possibility is that the blend is the lens.  Recall
that we concluded that this was against the ``balance of evidence''
but was still possible given the systematic astrometric errors induced
by unresolved stars.  A high resolution image taken ``immediately''
(i.e., during 2019) could largely resolve this question.  Recall that
the astrometric position of the lens relative to the KMT template is
known with an error of only 5 mas.  The astrometric error of a  high
resolution image would almost certainly be even smaller.  
Hence, the main uncertainty
in identifying these two positions will simply be the displacement
of the blend star relative to the field stars due to its unknown
proper motion.  Although this
will already be of order 5 mas in 2019, and will continue to grow,
in principle it can be corrected based on a subsequent high-resolution
image.  If this imaging shows a substantial offset, then the
blend is definitely not the lens.  And if the offset is within, say,
10 mas, then most likely it is lens.  In particular, such a small offset
would decisively exclude the blend as a companion to the lens because
such a close companion would have given rise to a huge microlensing
signal near the peak of the event.  And it would also virtually exclude
the scenario that the blend is an ambient star because the probability for an $I<19$
star to lie within 10 mas of a pre-defined location is $<10^{-3}$.  Hence,
the only remaining possible identifications for the blend
would be the lens itself or a companion to the source.

Once the results of this early high-resolution imaging are known,
then the next steps can be decided.  Consequently, we do not
attempt to chart these in excessive detail.  However, we note
that if the blended light proves to be the lens, then it will
be possible to probe the planetary system using radial-velocity (RV)
measurements with next generation telescopes.  For example, in the
$(s_2<1,u_0<0)$ solution discussed in Section~\ref{sec:blend}, 
the projected separation of the ``Jupiter''
would be $\sim 1.7\,\au$, while the mass ratio would be about 
$q_2\sim 2.2\times 10^{-3}$, implying a potential RV amplitude as
high as $\sim 50\,{\rm m\,s^{-1}}$, with a period as short as $P=2.2\,{\rm yr}$.
Such measurements on an $I<19$ star are plausibly feasible for 30 m telescopes.
Of course, in the wide solutions, the maximum possible
amplitude would be $\sim 20\,{\rm m\,s^{-1}}$, and the period would be of order
12 times longer.  Nevertheless (again, if the blend were shown to
be the lens), there would be at least some prospect for probing
this system.


\acknowledgments 
Work by AG was supported by AST-1516842 from the US NSF.
IGS and AG were supported by JPL grant 1500811.
AG received support from the European  Research  Council  under  the  European  Union’s Seventh Framework Programme (FP 7) ERC Grant Agreement n. [321035]
Work by C.H. was supported by the grant (2017R1A4A1015178)
of National Research Foundation of Korea.
This research has made use of the KMTNet system operated by the Korea
Astronomy and Space Science Institute (KASI) and the data were obtained at
three host sites of CTIO in Chile, SAAO in South Africa, and SSO in
Australia.
The OGLE project has received funding from the National Science Centre,
Poland, grant MAESTRO 2014/14/A/ST9/00121 to AU.
This research uses data obtained through the Telescope Access Program
(TAP), which has been funded by the National Astronomical
Observatories of China, the Chinese Academy of Sciences (the Strategic
Priority Research Program “The Emergence of Cosmological Structures”
Grant No. XDB09000000), and the Special Fund for Astronomy from the
Ministry of Finance. This work was partly supported by the National
Science Foundation of China (grant No. 11333003, 11390372 to
SM). Partly based on observations obtained with MegaPrime/MegaCam, a
joint project of CFHT and CEA/DAPNIA, at the Canada–France–Hawaii
Telescope (CFHT) which is operated by the National Research Council
(NRC) of Canada, the Institut National des Science de l’Univers of the
Centre National de la Recherche Scientifique (CNRS) of France, and the
University of Hawaii. The authors wish to recognize and acknowledge
the very significant cultural role and reverence that the summit of
Mauna Kea has always had within the indigenous Hawaiian community. We
are most fortunate to have the opportunity to conduct observations
from this mountain.

This work was performed
in part under contract with the California Institute of Technology
(Caltech)/Jet Propulsion Laboratory (JPL) funded by NASA through the
Sagan Fellowship Program executed by the NASA Exoplanet Science
Institute. MTP was supported by NASA grants NNX14AF63G and NNG16PJ32C,
as well as the Thomas Jefferson Chair for Discovery and Space
Exploration.

\appendix
\section{Decision to Search for 3L1S Solutions}
\label{sec:append3L1S}

As briefly telegraphed in Section~\ref{sec:2L1Sprob}, our decision
to carry out the 3L1S investigation was prompted by ``accidental''
developments in the course of the 2L1S investigation, i.e., 
apparent ``problems'' in the 2L1S solution that were all eventually resolved.
Such detours are normal in any relatively complex scientific study, but
the ``lab notebook details'' describing them are usually omitted
from journal papers describing the work, in order to avoid burdening
the reader with the tribulations of the authors.
In the present case, however, these ``accidental'' developments
led the investigation in an unexpected new direction, which could
have important implications for both the microlensing event described
here and for future events in its class.  For completeness, we 
therefore give a brief account of how we stumbled upon this direction.

As discussed in Section~\ref{sec:2L1S}, parallax and orbital motion
were introduced simultaneously, but the orbital-motion only improved
the fit by attaching a caustic feature to the baseline light curve.
However, this was not recognized immediately, and the moderately significant
differences between the ``standard'' and the ``higher-order'' solutions
were attributed mainly to parallax.  In particular, the biggest change
was a much smaller value of $\rho$ (so larger $\theta_\e$).
When combined with the value of $\pi_\e$ in that higher-order 
solution (about half the values in Table~\ref{tab:2L1S}), this yielded a lens
mass and distance (Equation~(\ref{eqn:massdist})) that
predicted a lens flux that was strongly inconsistent with the upper
limits derived in Section~\ref{sec:baseline}.

We were led to investigate 3L1S solutions because they could
solve this contradiction in one of two ways.  First, if the
host were actually a close binary, it could keep the same total
mass but generate much less light because the total mass would be divided
into two stars.  Indeed this effect played a crucial role in the solution
of OGLE-2007-BLG-349 \citep{ob07349}.  Second, the tidal shear due
to a wide binary (or corresponding quadrupole distortion of a close binary,
\citealt{dominik99}), could have affected the $\pi_{\e,N}$ measurement,
decreasing its amplitude relative to the true value, and so 
incorrectly raising the inferred mass.

However, the net result was mostly contrary to the
expectations in the previous paragraph.  Most importantly, the
third body turned out to have $q\ll 1$, so in the close solution
it did not play any role in reducing the total lens light.  Second,
the lens mass was actually reduced mostly by increasing $\rho$
(so decreasing $\theta_\e$) back near the level of the ``standard''
2L1S solution.  In addition, the mass was further reduced by
an increase in $|\pi_{\e,N}|$, which was in accord with our 
naively reasoned expectations.

Because of these puzzling results (particularly the fact that the
3L1S ``higher order'' solution looked more like the 2L1S ``standard'' 
solution than it looked like the 2L1S ``higher order'' solution),
we undertook an investigation
that normally would have been made only at a later stage: examining cumulative
$\Delta\chi^2$ plots between different solutions in order to locate
the times of observations and the individual observatories that were
contributing the most to distinguishing between solutions.  It was
at this point that we discovered that the orbital-motion parameters
were leading to spurious solutions through coupling to noise in the
baseline.  After removing these parameters, we obtained the solutions
that are discussed in Section~\ref{sec:2L1S}.  

\begin{deluxetable}{lccc}
\tablecolumns{4} \tablewidth{0pc} \tablecaption{\textsc{Best-Fit solutions for 2L1S models}} \tablehead{ \colhead{ } & \colhead{ }&
\multicolumn{2}{c}{Parallax models}\\
\cline{3-4} \colhead{Parameters } & \colhead{Standard}&
\colhead{$u_0>0$}&\colhead{$u_0<0$}} \startdata
  $\chi^2/\rm{dof}$               &9240.291/9124         &9169.300/9122         &9169.877/9122        \\
  $t_0$ $(\rm{HJD}^{\prime})$     &8219.579 $\pm$ 0.004  &8219.579 $\pm$ 0.004  &8219.579 $\pm$ 0.004 \\
  $u_0$ $(10^{-3})$               &9.341 $\pm$ 0.371     &8.295 $\pm$ 0.353     &-7.873 $\pm$ 0.369   \\
  $t_{\rm E}$ $(\rm{days})$       &118.327 $\pm$ 4.709   &133.269 $\pm$ 5.849   &140.226 $\pm$ 5.962  \\
  $s$                             &1.014 $\pm$ 0.0005    &1.012 $\pm$ 0.001     &1.013 $\pm$ 0.001    \\
  $q$ $(10^{-4})$                 &1.389 $\pm$ 0.053     &1.015 $\pm$ 0.057     &0.962 $\pm$ 0.056    \\
  $\alpha$ $(\rm{rad})$           &-0.431 $\pm$ 0.002    &-0.465 $\pm$ 0.005    &0.465 $\pm$ 0.005    \\
  $\rho$ $(10^{-4})$              &3.154 $\pm$ 0.099     &2.734 $\pm$ 0.105     &2.680 $\pm$ 0.101    \\
  $\pi_{\rm{E},\it{N}}$           &-                     &-0.785 $\pm$ 0.105    &0.765 $\pm$ 0.097    \\
  $\pi_{\rm{E},\it{E}}$           &-                     &-0.104 $\pm$ 0.018    &-0.054 $\pm$ 0.015   \\
  $f_S$                           &0.0201 $\pm$ 0.0008   &0.0181 $\pm$ 0.0008   &0.0171 $\pm$ 0.0008  \\
  $f_B$                           &0.7764 $\pm$ 0.0008   &0.7787 $\pm$ 0.0008   &0.7797 $\pm$ 0.0008  \\
  $t_*$ $(\rm{days})$             &0.037 $\pm$ 0.001     &0.036 $\pm$ 0.001     &0.038 $\pm$ 0.002    \\
\enddata
\tablecomments{$t_*\equiv \rho t_\e$ is not an independent quantity.} 
\label{tab:2L1S}
\end{deluxetable}

\begin{deluxetable}{lccccc}
\tablecolumns{6} \tablewidth{0pc} \tablecaption{\textsc{Best-fit solutions for 3L1S models}} \tablehead{ \colhead{} &
\multicolumn{2}{c}{Close} & \colhead{} &
\multicolumn{2}{c}{Wide}\\
\cline{2-3} \cline{5-6} \colhead{Parameters} & \colhead{$u_0>0$} &
\colhead{$u_0<0$} & \colhead{} & \colhead{$u_0>0$} &
\colhead{$u_0<0$}} \startdata
  $\chi^2/\rm{dof}$               &9088.321/9119         &9092.882/9119        & &9090.243/9119         &9093.705/9119        \\
  $t_0$ $(\rm{HJD}^{\prime})$     &8219.590 $\pm$ 0.009  &8219.571 $\pm$ 0.008 & &8219.479 $\pm$ 0.028  &8219.493 $\pm$ 0.028 \\
  $u_0$ $(10^{-3})$               &8.228 $\pm$ 0.613     &-8.893 $\pm$ 0.645   & &7.133 $\pm$ 0.503     &-7.428 $\pm$ 0.709   \\
  $t_{\rm E}$ $(\rm{days})$       &139.317 $\pm$ 9.735   &128.530 $\pm$ 9.344  & &151.039 $\pm$ 8.858   &145.361 $\pm$ 11.011 \\
  $s_1$                           &1.013 $\pm$ 0.001     &1.013 $\pm$ 0.001    & &1.011 $\pm$ 0.001     &1.011 $\pm$ 0.001    \\
  $q_1$ $(10^{-4})$               &0.975 $\pm$ 0.120     &1.187 $\pm$ 0.133    & &0.927 $\pm$ 0.100     &0.963 $\pm$ 0.168    \\
  $\alpha$ $(\rm{rad})$           &-0.477 $\pm$ 0.007    &0.475 $\pm$ 0.007    & &-0.478 $\pm$ 0.007    &0.474 $\pm$ 0.009    \\
  $\rho$ $(10^{-4})$              &2.793 $\pm$ 0.240     &3.310 $\pm$ 0.255    & &2.905 $\pm$ 0.198     &2.884 $\pm$ 0.308    \\
  $s_2$                           &0.364 $\pm$ 0.043     &0.406 $\pm$ 0.041    & &2.656 $\pm$ 0.375     &2.634 $\pm$ 0.396    \\
  $q_2$ $(10^{-3})$               &3.081 $\pm$ 0.871     &2.231 $\pm$ 0.712    & &2.456 $\pm$ 1.059     &2.208 $\pm$ 1.022    \\
  $\psi$ $(\rm{rad})$             &-0.036 $\pm$ 0.047    &0.144 $\pm$ 0.042    & &-0.086 $\pm$ 0.047    &0.084 $\pm$ 0.055    \\
  $\pi_{\rm{E},\it{N}}$           &-0.675 $\pm$ 0.129    &0.473 $\pm$ 0.157    & &-0.630 $\pm$ 0.100    &0.625 $\pm$ 0.227    \\
  $\pi_{\rm{E},\it{E}}$           &-0.105 $\pm$ 0.018    &-0.041 $\pm$ 0.016   & &-0.087 $\pm$ 0.016    &-0.038 $\pm$ 0.015   \\
  $f_S$                           &0.0175 $\pm$ 0.0013   &0.0189 $\pm$ 0.0013  & &0.0161 $\pm$ 0.0011   &0.0168 $\pm$ 0.0015  \\
  $f_B$                           &0.7790 $\pm$ 0.0013   &0.7776 $\pm$ 0.0013  & &0.7804 $\pm$ 0.0011   &0.7799 $\pm$ 0.0015  \\
  $t_*$ $(\rm{days})$             &0.039 $\pm$ 0.002     &0.043 $\pm$ 0.002    & &0.044 $\pm$ 0.002     &0.042 $\pm$ 0.002    \\
\enddata
\tablecomments{$t_*\equiv \rho t_\e$ is not an independent quantity.} 
\label{tab:3L1S}
\end{deluxetable}

\begin{deluxetable}{lcccc}
\tablecolumns{7} \tablewidth{0pc}\tablecaption{\textsc{Best-fit
solutions for 2L2S models}} \tablehead{ \colhead{ } & \colhead{ } &
 \multicolumn{2}{c}{Parallax}
\\\cline{3-4}
\colhead{Parameters} & \colhead{Standard}& \colhead{$u_0>0$}&
\colhead{$u_0<0$} }
\startdata
  $\chi^2/\rm{dof}$                   &9116.315/9121         &9104.215/9119        &9106.616/9119     \\
  $t_{0,1}$ $(\rm{HJD}^{\prime})$     &8219.598 $\pm$ 0.005  &8219.594 $\pm$ 0.005 &8219.594 $\pm$ 0.005 \\
  $u_{0,1}$ $(10^{-3})$               &7.832 $\pm$ 0.429     &5.926 $\pm$ 0.392    &-5.557 $\pm$ 0.438 \\
  $t_{\rm E}$ $(\rm{days})$           &131.607 $\pm$ 6.148   &173.809 $\pm$ 8.555  &182.493 $\pm$ 11.562 \\
  $s$                                 &1.014 $\pm$ 0.0006    &1.013 $\pm$ 0.0006   &1.014 $\pm$ 0.0006 \\
  $q$ $(10^{-4})$                     &1.110 $\pm$ 0.060     &0.705 $\pm$ 0.056    &0.660 $\pm$ 0.064   \\
  $\alpha$ $(\rm{rad})$               &-0.403 $\pm$ 0.004    &-0.436 $\pm$ 0.007   &0.431 $\pm$ 0.007  \\
  $\rho_1$ $(10^{-4})$                &2.747 $\pm$ 0.112     &2.149 $\pm$ 0.122    &1.986 $\pm$ 0.135   \\
  $t_{0,2}$ $(\rm{HJD}^{\prime})$     &8217.655 $\pm$ 0.598  &8219.763 $\pm$ 0.575 &8220.006 $\pm$ 0.342 \\
  $u_{0,2}$                           &0.048 $\pm$ 0.007     &0.026 $\pm$ 0.006    &-0.023 $\pm$ 0.004 \\
  $qF,I$                              &0.137 $\pm$ 0.017     &0.119 $\pm$ 0.018    &0.130 $\pm$ 0.017 \\
  $\pi_{\rm{E},\it{N}}$               &-                     &-0.557 $\pm$ 0.095   &0.527 $\pm$ 0.098    \\
  $\pi_{\rm{E},\it{E}}$               &-                     &-0.072 $\pm$ 0.020   &-0.041 $\pm$ 0.015  \\
  $f_S$                               &0.019 $\pm$ 0.0011    &0.014 $\pm$ 0.0009   &0.013 $\pm$ 0.0010 \\ 
  $f_B$                               &0.777 $\pm$ 0.0010    &0.782 $\pm$ 0.0009   &0.783 $\pm$ 0.0010 \\
  $t_*$ $(\rm{days})$                 &0.036 $\pm$ 0.002     &0.037 $\pm$ 0.002    &0.036 $\pm$ 0.002 \\
  \enddata
\tablecomments{$t_*\equiv \rho t_\e$ is not an independent quantity.} 
\label{tab:2L2S}
\end{deluxetable}

\begin{deluxetable}{lcc}
\tablecolumns{3} \tablewidth{0pc} \tablecaption{\textsc{Physical
parameters for 2L1S models}} \tablehead{\colhead{Quantity} &
\colhead{$u_0>0$} & \colhead{$u_0<0$}} \startdata
  $M_{\rm lens}$ $[M_\sun]$             &$0.195_{-0.020}^{+0.022}$  &$0.203_{-0.020}^{+0.022}$   \\
  $M_{\rm planet}$ $[M_\earth]$         &$6.228_{-0.881}^{+0.905}$  &$6.554_{-0.808}^{+0.905}$   \\
  $a_{\bot}$ [au]                       &$1.064_{-0.115}^{+0.119}$  &$1.103_{-0.107}^{+0.118}$   \\
  $a_{\bot}/a_{\rm snow}$               &$2.015_{-0.052}^{+0.051}$  &$2.011_{-0.051}^{+0.051}$   \\
  ${\it D_L}$ [kpc]                     &$0.773_{-0.099}^{+0.102}$  &$0.804_{-0.092}^{+0.100}$   \\
  $\mu_{\rm geo}$ [mas/yr]              &$3.587_{-0.170}^{+0.198}$  &$3.594_{-0.170}^{+0.209}$   \\
  $\mu_{{\rm hel},N}$ [mas/yr]          &$-2.880_{-0.177}^{+0.154}$ &$4.230_{-0.221}^{+0.273}$   \\
  $\mu_{{\rm hel},E}$ [mas/yr]          &$2.043_{-0.341}^{+0.412}$  &$2.196_{-0.317}^{+0.360}$   \\
  $\tilde{\rm v}_{{\rm lsr},l}$ [km/s]  &$5.872_{-1.814}^{+1.740}$  &$32.279_{-1.386}^{+1.540}$  \\
  $\tilde{\rm v}_{{\rm lsr},b}$ [km/s]  &$-6.049_{-0.705}^{+0.663}$ &$7.809_{-0.984}^{+1.148}$   \\
\enddata
\tablecomments{The error bars for $a_{\bot}/a_{\rm snow}$ are calculated
under the definition $a_{\rm snow}\equiv 2.7\,\au(M/M_\odot)$.} 
\label{tab:2L1Sphys}
\end{deluxetable}

\begin{deluxetable}{lccccc}
\tablecolumns{6} \tablewidth{0pc} \tablecaption{\textsc{Physical
parameters for 3L1S models}} \tablehead{ \colhead{} &
\multicolumn{2}{c}{Close} & \colhead{} &
\multicolumn{2}{c}{Wide}\\
\cline{2-3} \cline{5-6} \colhead{Quantity} & \colhead{$u_0>0$} &
\colhead{$u_0<0$} & \colhead{} & \colhead{$u_0>0$} &
\colhead{$u_0<0$}} \startdata
  $M_{\rm lens}$ $[M_\sun]$             &$0.238_{-0.029}^{+0.056}$  &$0.254_{-0.043}^{+0.110}$   & &$0.242_{-0.026}^{+0.041}$  &$0.254_{-0.034}^{+0.076}$  \\
  $M_{\rm planet,1}$ $[M_\earth]$       &$8.041_{-1.274}^{+2.877}$  &$9.505_{-2.224}^{+5.645}$   & &$7.697_{-1.108}^{+1.894}$  &$8.309_{-1.501}^{+3.995}$  \\
  $M_{\rm planet,2}$ $[M_J]$            &$0.679_{-0.188}^{+0.309}$  &$0.613_{-0.197}^{+0.346}$   & &$0.746_{-0.239}^{+0.341}$  &$0.743_{-0.241}^{+0.440}$  \\
  $a_{1\bot}$ [au]                      &$1.363_{-0.156}^{+0.309}$  &$1.484_{-0.248}^{+0.548}$   & &$1.364_{-0.136}^{+0.225}$  &$1.434_{-0.181}^{+0.402}$  \\
  $a_{2\bot}$ [au]                      &$0.506_{-0.087}^{+0.146}$  &$0.599_{-0.126}^{+0.244}$   & &$3.819_{-0.583}^{+0.817}$  &$3.963_{-0.672}^{+1.080}$  \\
  $a_{1\bot}/a_{\rm snow}$          &$2.113_{-0.084}^{+0.080}$  &$2.130_{-0.102}^{+0.086}$   & &$2.083_{-0.079}^{+0.078}$  &$2.075_{-0.102}^{+0.088}$  \\
  $a_{2\bot}/a_{\rm snow}$          &$0.784_{-0.102}^{+0.104}$  &$0.851_{-0.104}^{+0.099}$   & &$5.724_{-0.560}^{+0.720}$  &$5.536_{-0.621}^{+0.740}$  \\
  ${\it D_L}$ [kpc]                     &$1.091_{-0.153}^{+0.307}$  &$1.239_{-0.255}^{+0.529}$   & &$1.077_{-0.133}^{+0.223}$  &$1.150_{-0.184}^{+0.398}$  \\
  $\mu_{\rm geo}$ [mas/yr]              &$3.263_{-0.191}^{+0.216}$  &$3.314_{-0.190}^{+0.217}$   & &$3.162_{-0.180}^{+0.203}$  &$3.178_{-0.170}^{+0.205}$  \\
  $\mu_{{\rm hel},N}$ [mas/yr]          &$-2.785_{-0.203}^{+0.181}$ &$3.670_{-0.226}^{+0.274}$   & &$-2.686_{-0.186}^{+0.186}$ &$3.569_{-0.234}^{+0.244}$  \\
  $\mu_{{\rm hel},E}$ [mas/yr]          &$1.220_{-0.492}^{+0.342}$  &$1.228_{-0.658}^{+0.474}$   & &$1.286_{-0.376}^{+0.298}$  &$1.375_{-0.570}^{+0.375}$  \\
  $\tilde{\rm v}_{{\rm lsr},l}$ [km/s]  &$1.167_{-6.400}^{+2.800}$  &$38.570_{-4.299}^{+10.095}$ & &$1.957_{-4.214}^{+2.400}$  &$36.000_{-2.838}^{+6.500}$ \\
  $\tilde{\rm v}_{{\rm lsr},b}$ [km/s]  &$-7.633_{-2.033}^{+1.033}$ &$12.255_{-3.020}^{+8.475}$  & &$-7.486_{-1.500}^{+0.914}$ &$10.375_{-2.000}^{+5.250}$ \\
\enddata
\tablecomments{The error bars for $a_{\bot}/a_{\rm snow}$ 
are calculated under the definition $a_{\rm snow}\equiv 2.7\,\au(M/M_\odot)$.} 
\label{tab:3L1Sphys}
\end{deluxetable}

\begin{deluxetable}{lccc}
\tablecolumns{6} \tablewidth{0pc} \tablecaption{\textsc{Physical
parameters for 2L2S models}} \tablehead{ \colhead{} &
\multicolumn{2}{c}{Parallax} \\
\cline{2-3} \colhead{Quantity} & \colhead{$u_0>0$} &
\colhead{$u_0<0$}} \startdata
  $M_{\rm lens}$ $[M_\sun]$             &$0.324_{-0.041}^{+0.053}$  &$0.342_{-0.046}^{+0.055}$   \\
  $M_{\rm planet}$ $[M_\earth]$         &$8.389_{-1.110}^{+1.629}$  &$8.058_{-1.102}^{+1.462}$     \\
  $a_{\bot}$ [au]                       &$1.536_{-0.174}^{+0.226}$  &$1.598_{-0.193}^{+0.219}$    \\
  $a_{\bot}/a_{\rm snow}$ [au]          &$1.757_{-0.067}^{+0.066}$  &$1.726_{-0.067}^{+0.064}$     \\
  ${\it D_L}$ [kpc]                     &$1.008_{-0.129}^{+0.167}$  &$1.034_{-0.142}^{+0.153}$     \\
  $\mu_{\rm geo}$ [mas/yr]              &$3.439_{-0.190}^{+0.250}$  &$3.219_{-0.204}^{+0.265}$     \\
  $\mu_{{\rm hel},N}$ [mas/yr]          &$-2.911_{-0.215}^{+0.177}$ &$3.689_{-0.249}^{+0.318}$     \\
  $\mu_{{\rm hel},E}$ [mas/yr]          &$1.344_{-0.335}^{+0.317}$  &$1.535_{-0.304}^{+0.342}$     \\
  $\tilde{\rm v}_{{\rm lsr},l}$ [km/s]  &$1.782_{-3.245}^{+2.166}$  &$34.370_{-1.840}^{+2.358}$    \\
  $\tilde{\rm v}_{{\rm lsr},b}$ [km/s]  &$-7.300_{-1.199}^{+0.887}$ &$9.765_{-1.424}^{+1.925}$     \\
\enddata
\label{tab:2L2Sphys}
\end{deluxetable}

\begin{deluxetable}{lccc}
\tablecolumns{4} \tablewidth{0pc} \tablecaption{\textsc{3L1S
parallax-only models for fixed $\pi_{\rm{E},\it{N}}$}} \tablehead{
\colhead{} & \colhead{Close} & \colhead{Wide}\\
\colhead{Parameters} & \colhead{$\pi_{\rm{E},\it{N}}=0.12$} &
\colhead{$\pi_{\rm{E},\it{N}}=0.12$}} \startdata
  $\chi^2/\rm{dof}$               &9095.359/9120        &9095.939/9120        \\
  $t_0$ $(\rm{HJD}^{\prime})$     &8219.571 $\pm$ 0.007 &8219.494 $\pm$ 0.018 \\
  $u_0$ $(10^{-3})$               &-9.426 $\pm$ 0.586   &-9.240 $\pm$ 0.488   \\
  $t_{\rm E}$ $(\rm{days})$       &121.167 $\pm$ 7.127  &118.429 $\pm$ 5.344  \\
  $s_1$                           &1.014 $\pm$ 0.001    &1.011 $\pm$ 0.001    \\
  $q_1$ $(10^{-4})$               &1.371 $\pm$ 0.099    &1.401 $\pm$ 0.083    \\
  $\alpha$ $(\rm{rad})$           &0.462 $\pm$ 0.005    &0.463 $\pm$ 0.005    \\
  $\rho$ $(10^{-4})$              &3.618 $\pm$ 0.186    &3.658 $\pm$ 0.174    \\
  $s_2$                           &0.425 $\pm$ 0.033    &2.251 $\pm$ 0.636    \\
  $q_2$ $(10^{-3})$               &2.179 $\pm$ 0.580    &2.311 $\pm$ 0.638    \\
  $\psi$ $(\rm{rad})$             &-0.160 $\pm$ 0.036   &0.141 $\pm$ 0.034    \\
  $\pi_{\rm{E},\it{E}}$           &-0.044 $\pm$ 0.015   &-0.056 $\pm$ 0.015   \\
  $f_S$                           &0.0200 $\pm$ 0.0013  &0.0205 $\pm$ 0.0011  \\
  $f_B$                           &0.7764 $\pm$ 0.0012  &0.7759 $\pm$ 0.0010  \\
  $t_*$ $(\rm{days})$             &0.044 $\pm$ 0.002    &0.043 $\pm$ 0.002    \\
\enddata
\tablecomments{} \label{tab:forced}
\end{deluxetable}

\end{document}